\documentclass[aps,prd,reprint,superscriptaddress,floatfix]{revtex4-2}

\usepackage{graphicx}
\usepackage{dcolumn}
\usepackage{amsmath}
\usepackage{bm}
\usepackage{hyperref}
\usepackage{color}
\usepackage{float}

\newcommand{\chimera}{\textsc{Chimera}}
\newcommand{\msun}{\ensuremath{M_\odot}}

\begin{document}

\title{\textbf{Core Collapse Supernova Gravitational Wave Sourcing and Characterization \\ based on Three-Dimensional Models} 
}

\author{R. Daniel Murphy}
\affiliation{
Department of Physics and Astronomy, University of Tennessee, Knoxville, Knoxville, TN 37996-1200, USA
}
\email{Contact author:rmurph16@vols.utk.edu}
\author{Anthony Mezzacappa}
\affiliation{
Department of Physics and Astronomy, University of Tennessee, Knoxville, Knoxville, TN 37996-1200, USA
}
\author{Eric J. Lentz}
\affiliation{
Department of Physics and Astronomy, University of Tennessee, Knoxville, Knoxville, TN 37996-1200, USA
}
\affiliation{
Physics Division, Oak Ridge National Laboratory, P.O. Box 2008, Oak Ridge, TN 37831-6354, USA
}
\author{Pedro Marronetti}
\affiliation{
Physics Division, National Science Foundation, Alexandria, Virginia 22314, USA
}

\date{\today}

\begin{abstract}
We present for the first time an analysis of high-frequency gravitational wave emission from proto-neutron stars (PNS) in core collapse supernovae (CCSNe) that combines spatial decomposition and modal decomposition to both source and characterize the emission using three-dimensional CCSN simulations. Our analysis is based on three-dimensional CCSN simulations initiated from a Solar-metallicity 15~\msun\, progenitor and a zero-metallicity 25~\msun\ progenitor. 
We decompose the gravitational wave strains into five spatial regions, and find that strains are initially largest in the PNS surface layers from accretion and later largest from the Ledoux convective and convective overshoot regions deep within the PNS. 
We compute the fractional gravitational wave luminosity as a function of enclosed radius and observe that the majority of the luminosity moves from the PNS surface at $\sim$100 ms postbounce to deep within the PNS at later times. 
Using a self-consistent perturbative analysis, we investigate the evolution of the quadrupolar, non-radial, quasi-normal oscillation modes of the PNS. We find that the frequency of the evolving high-frequency component of the gravitational wave signal is well matched to the eigenfrequency evolution of the ${}^2g_2$-mode initially, then the ${}^2g_1$-mode, and finally the ${}^2f$-mode, as labeled by the Generalized Cowling Nomenclature.
We show that the ${}^2g$-modes emit most of their power in gravitational waves initially from the PNS surface region, but within a few 100 ms after bounce, it is the convective overshoot region of the PNS, just above the region of sustained Ledoux convection, that emits the most gravitational wave power for the ${}^2g_1$-mode. Eventually, the ${}^2f$-mode is the dominant mode producing gravitational waves, and these waves are emitted primarily from the convective overshoot region.
Thus, with three interconnected analyses, we show that, while the gravitational wave emission is global, stemming from multiple regions in and around the PNS, it nonetheless remains possible to source the dominant contributions to it. 
We find that the source of the high-frequency gravitational wave emission from the PNS in CCSNe is more complex than assessed by other methods, as well as time dependent, first emitted mainly by ${}^2g$-modes driven by accretion onto the PNS and later emitted by the ${}^2f$-mode driven by sustained Ledoux convection.
 \end{abstract}

\maketitle

\section{Introduction}
\label{sec:introduction}
Core collapse supernova theory has advanced significantly in the past decade. Sophisticated three-dimensional explosion models have been painstakingly developed and published during this time by groups around the world. In particular, the efficacy of neutrino shock reheating to power the explosions, with the aid of fluid and shock instabilities, has been demonstrated. And the spectrum of possible explosion mechanisms---specifically, neutrino-driven versus magnetohydrodynamically-driven explosions---has been more clearly defined. For reviews of the progress made in multidimensional core collapse supernova modeling that has brought us to the current state of the art, we refer the reader to
\citet{Mezz05},
\citet{KoSaTa06},
\citet{JaLaMa07},
\citet{JaHaHu12},
\citet{Burr13},
\citet{JaMeSu16},
\citet{Mull16},
\citet{MeEnMe20},
\citet{Mull20},
\citet{BuVa21},
and
\citet{YaNaAk24}.

The excitement over progress made in modeling core collapse supernovae is in part due to the fact that we are now in a position to make sophisticated predictions for multimessenger emissions---in particular, gravitational wave emissions---from these events. The new window on the Universe that was opened with the first detection of gravitational waves from a binary black hole merger \cite{GW151226}, which also occurred within the past decade, and the fact that gravitational waves will emerge unimpeded from the deepest regions of these supernovae and will bear the imprints of the components of the central engine driving them (e.g., see \citet{MeZa24} for a complete discussion of the association between components of the explosion mechanism and components of the gravitational wave emission), make this a particularly exciting time to model core collapse supernovae. A Galactic or near-extra-Galactic event will enable validation of supernova models, as well as parameter estimation of important physical properties of these astrophysical systems, such as macroscopic properties of the remnant PNS (e.g., masses and radii), and perhaps microscopic properties such as the nuclear equations of state (e.g., see \cite{MuCaMe24} and references therein).

The full benefits of a core collapse supernova gravitational wave detection cannot be enjoyed unless the gravitational wave emission is both sourced and characterized. With regard to the former, several groups have endeavored to decompose the emission spatially \cite{AnMuMu17,MeMaLa20,MeMaLa23}. In particular, in \citet{MeMaLa20,MeMaLa23}, the gravitational wave emission was decomposed across five regions below the supernova shock wave. Moving outward in radius, they corresponded to (1) the region of sustained Ledoux convection deep within the PNS, (2) the region of convective overshoot lying directly above the convective region, (3) the surface layer of the PNS, (4) the neutrino-net-cooling layer above the PNS surface, and (5) the neutrino-net-heating, or gain, layer directly below the shock, wherein neutrino energy deposition occurs, powering the supernova. In the latter of these two studies, in the models considered, it was demonstrated that the gravitational wave emission is initially excited by accretion funnels impinging on the PNS surface and later by sustained Ledoux convection in the PNS. In the former case, the dominant emission---i.e., the largest strains---are initially found to be within the PNS surface layers just above and just below the PNS surface. In the latter case, the dominant emission was found to be within the convective overshoot layer within the PNS. The transition from one case to the other followed the natural evolution to explosion and the associated diminishment of accretion onto the PNS, as well as the development of sustained Ledoux convection within the PNS owing to the diffusion of neutrinos out of it on the characteristic timescale of $O(100)$ ms. Using indirect methods, other studies have concluded that gravitational waves are primarily excited by accretion onto the PNS, with contributions from deep within the PNS being minimal \cite{RaMoBu19,PoMu19,VaBuWa23}. These conclusions are not necessarily in tension with one another. As \citet{PoMu19} point out, different excitation mechanisms may be expected for different progenitor models, although the conclusions in \citet{MeMaLa23} did not vary across progenitors of different mass, metallicity, and internal structure. In any event, direct methods afford a way to address the differing conclusions drawn to date on this topic.

To characterize gravitational wave emission from core collapse supernovae, the community has turned to modal analyses/astroseismological studies---specifically, of the high-frequency emission from the PNS. (The studies cited above point to the gain layer as the origin of low-frequency emission resulting from the SASI and explosion.) Sotani and collaborators \cite{SoTa16,SoKuTa17,SoKuTa19,SoTa20,SoTa20a,SoTaTo21}, Torres-Forné and collaborators \cite{ToCePa18,ToCeOb19,ToCePa19}, Morozova and collaborators \cite{MoRaBu18,RaMoBu19}, and Westernacher-Schneider and collaborators \cite{WeOcSu19,West20} have pioneered PNS modal analyses based on the perturbation of a spherically symmetric background obtained by spherically averaging over multidimensional simulation data, all while assuming hydrostatic equilibrium. Obviously, for rotating stellar cores this presents a problem \cite{PoMu20,PoMuAg23,PoMu24}. In the models we consider here, our progenitors are nonrotating and, consequently, more amenable to the application of modal analysis methods as they have been applied in the past. With regard to hydrostatic equilibrium, \citet{ZhAnOc24} demonstrated that it does not apply strictly in the models they considered, though the degree to which it is violated is significantly less within the PNS and occurs largely, as expected, in the region encompassing the Ledoux convective layer. Despite the obvious shortcomings of such assumptions, the modal analyses produce results that are in good agreement with some results obtained directly from the underlying simulations. In particular, the predictions for the peak frequency of gravitational wave emission as a function of time after bounce track well the high-frequency feature (HFF) seen in all core collapse supernova simulation spectrograms, especially if the equations linearized as part of the modal analysis are the equations on which the simulations are based. \citet{WeOcSu19,West20} emphasized the need to use an appropriate treatment of gravity within modal analyses used on simulation data that arises from simulations in which general relativity is accounted for by using an ``effective potential'' to correct the Newtonian gravitational potential in an otherwise Newtonian treatment of gravity and hydrodynamics. 

The assignment of particular modes of oscillation of the PNS to the peak emission as a function of postbounce time is a bit less definitive. In two-dimensional analyses, some groups find the HFF is described well by either a low order ${}^2g$-mode \cite{ToCePa18,ToCePa19} or an ${}^2f$-mode \cite{AfKuCa23}, with the superscript denoting an $\ell=2$ quadrupolar mode. Others find the HFF begins as a low order ${}^2g$-mode and later becomes the fundamental ${}^2f$-mode \cite{MoRaBu18,SoTa20,RoRaCh23,ZhAnOc24,SoMuTa24}. Using three-dimensional simulation data for their modal analysis, \citet{RaMoBu19} and \citet{NaTaKo22} find agreement with the latter description of the HFF. Additionally, there are multiple classification methods, the most commonly used approach being the Cowling classification introduced by \citet{Cowl41}. \citet{RoRaCh23} have shown that the assignment of a particular mode to the emission can vary as the modal classification method varies. 

This becomes important when trying to fit the HFF to a polynomial that is a function of either the surface gravity of the source, or the mean density. It is well known from astroseismology that the peak frequency of $g$-modes depends on the surface gravity of the source, 
\begin{equation}
    f_{\rm peak}\sim \frac{GM}{R^2},
\label{eq:gmodepeak}
\end{equation}
where $G$ is Newton's gravitational constant, $M$ is the mass of the source, and $R$ is the radius of the source. Fits to the HFF using the surface gravity are found to work well in \cite{MuJaMa13,CeDeAl13,PaLiCo18,PoMu19,PoMu20}. The peak frequency of an $f$-mode depends on the square root of the mean density of the source, 
\begin{equation}
    f_{\rm peak}\sim\sqrt{\frac{M}{R^3}}.
\label{eq:fmodepeak}
\end{equation}
Thus, knowledge of the PNS oscillation mode producing the gravitational wave emission at a particular time after bounce---i.e., knowledge of the windows of the HFF in the gravitational wave spectrogram in which each mode dominates---is needed in order to fit the spectrogram to the appropriate source characteristic---i.e., surface gravity versus mean density---to generate so-called ``universal relations'' \cite{ToCeOb19,SoTaTo21,MoSuTa23,SoMuTa24}. \citet{ToCeOb19} use both, one to fit the ${}^2g$-mode component of the spectrogram and the other to fit the ${}^2f$-mode component. \citet{SoTaTo21}, on the other hand, attempt to fit the entire spectrogram with a single fit and found that the use of the mean density as the fundamental physical quantity on which the fit is based yielded the best results, and this was reaffirmed for different treatments of gravity in \citet{SoMuTa24}. \citet{MoSuTa23}, likewise attempt to fit both ${}^2g$- and ${}^2f$-modes from one-dimensional simulations with a single fit and find that coefficients can be chosen so that both the surface gravity and mean density based fits match the data well. They find that it is more important to include enough data to cover the late time behavior, at least out to 5 seconds postbounce, in order to determine the coefficients of the fit accurately. Given the fit, and the physical quantity underpinning it, the spectrogram can be used to determine the surface gravity or the mean density of the PNS, as a function of time after bounce \cite{BiMaTo21,PoMu22,BrBiOb23}. Further assumptions can isolate one of the quantities that enters into the expression for both the surface gravity and the mean density---i.e., the PNS mass, $M$, or its radius, $R$---in order to determine the other as a function of time after bounce, as the supernova unfolds.

In this paper, we improve upon, and augment, the sourcing of gravitational wave emission reported in \citet{MeMaLa20,MeMaLa23}, and use it for the models considered here. In addition, we conduct a modal analysis of these models based on a suitable treatment of gravity, as described by \citet{West20}, consistent with the pseudo-Newtonian treatment of gravity used in our \chimera\ simulation code, which was used to produce the simulation data on which our analyses are based. Our study aims to show that combining these direct methods of analysis affords a more complete understanding of the sourcing and characterization of gravitational wave emission than from the use of either ``tool'' individually. We conduct this analysis in the context of three-dimensional CCSN simulations, a significant advancement, and one explicitly called for, from the analysis conducted by \citet{Zha24} using two-dimensional CCSN simulations.

The paper is organized as follows: Section \ref{sec:models_and_methods} describes the \chimera\ CCSN simulation code and the progenitor models used and contains Subsection \ref{sub:GWS}, which gives a brief account of our method of gravitational wave extraction, as well as Subsection \ref{sub:Reg_Dec}, which describes our new regional decomposition. Subsection \ref{sub:PertAnalysis} derives the modal analysis of \citet{West20} in detail to account for all assumptions and approximations and describes our specific implementation. Section \ref{sec:results} presents our gravitational wave strains in our improved regional decomposition, their luminosities as a function of radius, the quasi-normal, quadrupolar oscillation mode solutions from our modal analysis, and the results of combining these methods of inquiry. Section \ref{sec:summary} discusses our results in the context of previous studies and summarizes our conclusions.

\section{Models and Methods}
\label{sec:models_and_methods}

Using an improved method of spatially decomposing the gravitational wave emission relative to what was reported in \citet{MeMaLa23}, together with an analysis of the PNS oscillation modes generating the gravitational wave emission, we investigated the gravitational wave emission from the collapse, and explosion, of the Solar-metallicity 15~\msun\ progenitor from \citet{WoHe07} and the zero-metallicity 25~\msun\ progenitor from \citet{HeWo10}, using three-dimensional core collapse supernova simulations performed with the \chimera\ code \cite{BrBlHi20}. These are part of the D-series simulations described in \citet{MeMaLa23}, with the former denoted as the D15 model and the later denoted as the D25 model. Both models were initially nonrotating. 

The \chimera\ code is a multiphysics code that combines Newtonian self-gravity with a monopole correction for general relativistic effects, multigroup flux-limited diffusion neutrino transport in the ray-by-ray approximation, Newtonian hydrodynamics, and a nuclear reaction network to simulate core collapse supernovae. Neutrino emission and absorption interactions included are electron capture on protons and nuclei, electron--positron annihilation and nucleon--nucleon bremsstrahlung and the corresponding inverse weak reactions. The neutrino scattering processes included are isoenergetic scattering on nuclei, and neutrino--electron and neutrino--nucleon scattering. The equation of state used in the D-series combines the EOS from \citet{LaSw91}  with a nuclear incompressibility of $K=220$ MeV at densities above $10^{11}$ g cm$^{-3}$ and the EOS from \citet{BaCoKa85} for densities below that. A 17-species alpha network computes nuclear burning in the outer regions with temperatures below 6.5~GK \cite{HiTh99}.
\chimera\ uses a two-component, overlapping `Yin-Yang' grid in spherical polar coordinates $(r,\theta,\phi)$, with the polar region of each component grid removed, at an angular resolution of one degree, and 720 radial zones. The innermost region of the star ($<8$ km) is evolved in spherical symmetry, with the radius of this region determined so that it always lies sufficiently below the convective region of the PNS. 

\subsection{Gravitational Wave Strains}
\label{sub:GWS}
We extract gravitational wave strains from the \chimera\, simulations using the methods described in section II.b of \citet{MeMaLa23}. Using the same procedure, we compute $N_{2m}$ given by
\begin{align}
    &N_{2m}=\frac{16\sqrt{3}\pi G}{15 c^4}\int^{2\pi}_{0}d\varphi'\int^\pi_0d\vartheta'\int^b_adr'r'^3\nonumber\\
           &\times\left[2\rho v^rY^*_{2m}\sin{\vartheta'}+\rho v^\vartheta\sin{\vartheta'}\frac{\partial}{\partial\varphi'}Y^*_{2m}+\rho v^\varphi\frac{\partial}{\partial \varphi'}Y^*_{2m}\right]\nonumber\\
           &-\frac{16\sqrt{3}\pi G}{15 c^4}\int_0^{2\pi}d\varphi'\int_0^\pi d\vartheta'Y^*_{2m}\sin{\vartheta'}\left(r^4_b\rho_bv_b^r-r^4_a\rho_av^r_a\right)\label{eq:N2m}
\end{align}
and defined by 
\begin{equation}
    N_{2m}\equiv\frac{G}{c^4}\frac{dI_{2m}}{dt}
\end{equation}
to determine the transverse-traceless gravitational wave strain as
\begin{equation}
    h^{TT}_{ij}=\frac{1}{r}\sum_{m=-2}^{+2}\frac{dN_{2m}}{dt}\left(t-\frac{r}{c}\right)f^{2m}_{ij},
\end{equation}
where $r$ is the radius of the observer, $m$ is the azimuthal component of the spherical harmonic, $t$ is the simulation time, and $f^{2m}_{ij}$ are the tensor spherical harmonics. We compute both polarizations of the strains, which relate to $h^{TT}_{ij}$ by
\begin{align}
    h_+&=\frac{h^{TT}_{\theta\theta}}{r^2},\\
    h_\times&=\frac{h^{TT}_{\theta\phi}}{r^2\rm{sin}\theta}.
\end{align}

\subsection{Regional Decomposition}
\label{sub:Reg_Dec}
The final term in Equation~\eqref{eq:N2m} is the surface term for the radial boundaries $r_a$ and $r_b$, which vanishes when the total strain is calculated, where $r_a=0$ and $r_b=\infty$, with $\rho_b=0$. However, as has been shown, this term cannot be neglected for calculations where $r_a\ne0$ and $r_b$ is finite \cite{EgZhSc21,MeMaLa23,Zha24}. If it is neglected, one loses the mathematical equivalence between $\ddot{I}_{2m}$ and $\dot{N}_{2m}$, and computations of the two quantities will diverge. In \citet{MeMaLa23}, this was known, but the regional decomposition implemented did not show agreement between methods in region 5 specifically, owing to the problems associated with using the shock as a regional boundary.

For this reason, we introduce a modification to the procedure applied in \cite{MeMaLa23}---specifically, a modification to the regional breakdown of the gravitational wave sources. Regions 1--4 are identical to our previous work. Region 1 is defined with an inner boundary at the innermost mean radius at which the convective mass flux is 5\% of its peak value, and the outer boundary is the outermost mean radius at which the convective mass flux is 5\% of its peak value. Region 2 extends from the outer boundary of region 1 to the mean density contour of $10^{12}$ g cm$^{-3}$. Region 3 is bounded below and above by the mean density contours of $10^{12}$ g cm$^{-3}$ and $10^{11}$ g cm$^{-3}$, respectively. Region 4 encompasses the net cooling region from the mean density contour of $10^{11}$ g cm$^{-3}$ to the mean gain radius, at which net neutrino-heating sets in. Previously, region 5 was defined to extend from the mean gain radius to the maximum shock radius.  However, because the contribution to gravitational wave emission outside the shock is minimal (the preshock flow is spherical given a spherical progenitor), extending region 5 to the edge of the computational domain circumvents challenges associated with using the shock as a regional boundary without compromising the determination of the gravitational wave emission itself. Thus, our new regional decomposition comprises the same regions 1--4, but region 5 extends from the gain radius to the edge of the computational grid rather than to the shock.

\subsection{Perturbative Analysis}
\label{sub:PertAnalysis}
For our modal analysis, we linearize the Newtonian hydrodynamic equations following methods in \citet{West20}. We also work in the spherical coordinate basis $\{(\partial_r)^i,(\partial_\theta)^i,(\partial_\phi)^i\}$. For a Newtonian perfect fluid, the equations relating density, $\rho$, pressure, $P$, and gravitational potential, $\Phi$, are given by
\begin{align}
\partial_t\rho+\nabla_i(\rho v^i)&=0, \label{eq:cont}\\
\partial_t(\rho v_i)+\nabla_j(\rho v^jv_i)+\partial_iP&=-\rho\partial_i\Phi,\label{eq:momentum}\\
\nabla^2\Phi&=4\pi\rho\label{eq:grav},
\end{align}
where $\partial_\mu$ denotes the partial derivative with respect to the $\mu$ component, $\nabla_i$ is the spatial covariant derivative, $v$ is the fluid velocity, Latin indices represent spatial components, and repeated indices are summed over. We also use geometrized units, with $G=c=1$.

We note that the results of \citet{West20} and \citet{ZhAnOc24} show good agreement between gravitational wave generation in pseudo-Newtonian core collapse supernova simulations and the eigenmodes computed from a purely Newtonian perturbative analysis. A true pseudo-Newtonian modal analysis would involve linearizing some equation of motion for the pseudo-Newtonian potential, instead of Equation~\eqref{eq:grav}. Considering a definition of the pseudo-Newtonian potential defined by Equation (53) in \citet{RaJa02}, we can see that introducing this equation would require the linearization of several additional terms and, ultimately, greatly complicate the system of equations needed to describe linear perturbations. The implementation of a pseudo-Newtonian perturbative method may be important to consider, and will be useful in quantifying the accuracy of the Newtonian approach of \citet{West20}, but is beyond the scope of this paper.

 Our final equations are identical to those of \citet{West20} and \citet{ZhAnOc24}. However, to facilitate reproducibility and use by others in any future studies, we wish to account for all assumptions and approximations made to arrive at these equations. To this end, we include in Subsections \ref{subsub:LE}--\ref{subsub:Solutions} a detailed derivation of the Newtonian modal analysis. In Subsection \ref{subsub:BC}, we describe and justify the boundary conditions we use in our numerical implementation, and in Subsection \ref{subsub:Scheme}, we outline how we solve the system to find the eigenfrequencies and eigenfunctions of the quasi-normal modes of oscillation of the PNS.

\subsubsection{Linearized Equations}
\label{subsub:LE}
We linearize about a spherically-averaged time slice of our three-dimensional simulation data such that thermodynamic variables are replaced by
\begin{equation}
    u\to u_0(r)+\delta u,
\end{equation}
where $u_0(r)$ is the spherically-averaged-background (equilibrium) quantity and $\delta u$ is the corresponding Eulerian perturbation. We note explicitly that because all background quantities are spherically averaged, they only have a radial dependence $\rho_0=\rho_0(r)$, $P_0=P_0(r)$, etc., while the perturbations contain, at this point, radial, polar, and azimuthal dependence. We assume that the background is in hydrostatic equilibrium, for which 
\begin{equation}
    v_0^i=0.
\end{equation}
Equation~\eqref{eq:momentum} for the background then becomes
\begin{align}
    \frac{\partial_rP_0}{\rho_0}=-\partial_r\Phi_0.\label{eq:HEC}
\end{align}
For simplicity, moving forward we drop the subscript, $0$, for background equilibrium quantities, as well as the explicit radial dependence of all quantities.

Linearizing with respect to the Eulerian perturbations $\delta U$, the continuity equation given by Equation~\eqref{eq:cont} is then written as
\begin{align}
\partial_t(\rho+\delta\rho)+\nabla_i\left[(\rho+\delta\rho)(v^i+\delta v^i)\right]&=0,
\end{align}
which yields
\begin{align}
\partial_t\delta\rho+\nabla(\rho\delta v^i+\delta\rho v^i+\delta\rho\delta v^i)&=-\partial_t\rho-\nabla_i(\rho v^i). 
\end{align}
The terms on the right-hand side sum to zero. They contain only background terms and satisfy Equation~\eqref{eq:cont}. Additionally, because $v^i=0$ (the background fluid is assumed to be in hydrostatic equilibrium) and we only keep linear perturbative terms, the second two terms in the covariant derivative above are eliminated. We then have
\begin{equation}
    \partial_t\delta\rho+\rho\nabla_i\delta v^i+\delta v^i\nabla_i\rho=0.
\end{equation}
Recall that for the flat 3-metric, $\gamma_{ij}$, the covariant derivative of a generic contravariant vector $V^i$ is given by $\nabla_i V^i=\partial_iV^i+V^i\partial_i\mathrm{ln}\sqrt{\gamma}$, where $\gamma$ is the determinant of the flat 3-metric and in spherical coordinates $\sqrt{\gamma}=r^2\mathrm{sin}\theta$. Then, 
\begin{equation}
\partial_t\delta\rho+\rho\partial_i\delta v^i+\rho\delta v^i \partial_i\mathrm{ln}\sqrt{\gamma}+\delta v^r\partial_r\rho=0,
\end{equation}
where we have also used the fact that $\rho=\rho(r)$. Therefore, $\nabla_i\rho=\partial_r\rho$ is the only nonzero derivative term. 

We can relate the Eulerian velocity perturbation to the Eulerian fluid element displacement vector $\xi^i$ by
\begin{equation}
    \delta v^i=\partial_t\xi^i+v^j\nabla_j\xi^i-\xi^j\nabla_jv^i,
\end{equation}
which reduces to $\delta v^i=\partial_t\xi^i$ when the background velocity is zero. Thus, 
\begin{equation}
\partial_t\delta\rho+\rho\partial_i\partial_t\xi^i+\rho\partial_t\xi^i \partial_i\mathrm{ln}\sqrt{\gamma}+\partial_t\xi^r\partial_r\rho=0.
\end{equation}
Guided by \citet{Poisson_Will_2014}, as done in \citet{West20}, we integrate this equation in time and set the integration constant to zero to obtain
\begin{equation}
\delta\rho+\rho\partial_i\xi^i+\rho\xi^i \partial_i\mathrm{ln}\sqrt{\gamma}+\xi^r\partial_r\rho=0.\label{eq:cont_final}
\end{equation}

The momentum equation given in Equation~\eqref{eq:momentum} is rewritten by making the same substitutions. The first term is given by
\begin{align}
    \partial_t(\rho v_i)&\rightarrow\partial_t\left[(\rho+\delta\rho)(v_i+\delta v_i)\right]\nonumber\\
&=\partial_t(\rho v_i)+\partial_t(\rho\delta v_i+\delta\rho v_i+\delta\rho\delta v_i)\nonumber\\
&\approx\partial_t(\rho v_i)+\partial_t(\rho\delta v_i)\nonumber\\
&=\partial_t(\rho v_i)+\partial_t(\rho\gamma_{ij}\delta v^j)\nonumber\\
&=\partial_t(\rho v_i)+\rho\gamma_{ij}\partial_t\delta v^j+\gamma_{ij}\delta v^j\partial_t\rho.
\end{align}
Given that the background fluid is in hydrostatic equilibrium with $v^i=0$, Equation~\eqref{eq:cont} results in $\partial_t\rho=0$, as well.
We then have
\begin{equation}
\partial_t(\rho v_i)\rightarrow\partial_t(\rho v_i)+\rho\gamma_{ij}\partial_t\delta v^j\label{eq:mom1}.
\end{equation}

The second term in Equation~\eqref{eq:momentum} becomes
\begin{align}
    \nabla_j(\rho v^jv_i)&\rightarrow\nabla\left[(\rho+\delta\rho)(v^j+\delta v^j)(v_i+\delta v_i)\right]\nonumber\\
    &=\nabla_j(\rho v^j v_i)+\nabla_j(\rho v^j \delta v_i+\rho \delta v^jv_i+\rho\delta v^j\delta v_i)\nonumber\\
    &\qquad+\nabla_j(\delta\rho v^jv_i+\delta\rho v^j \delta v_i+\delta\rho \delta v^jv_i\nonumber\\
    &\ \ \qquad\qquad+\delta\rho\delta v^j\delta v_i).\label{eq:step}
\end{align}
Neglecting terms containing higher-order perturbations and using the fact that the background is static, we see that the second and third terms in Equation~\eqref{eq:step} vanish. The second term in Equation~\eqref{eq:momentum} thus contributes no new terms:
\begin{equation}
    \nabla_j(\rho v^jv_i)\rightarrow\nabla_j(\rho v^j v_i).\label{eq:mom2}
\end{equation}

The final terms in the momentum equation transform straightforwardly to 
\begin{align}
    \partial_iP&\rightarrow\partial_i(P+\delta P)\nonumber\\
    &=\partial_iP+\partial_i\delta P\label{eq:mom3}
\end{align}
and
\begin{align}
-\rho\partial_i\Phi&\rightarrow-(\rho+\delta\rho)\partial_i(\Phi+\delta\Phi)\nonumber\\
&=-\rho\partial_i\Phi-\delta\rho\partial_i\Phi-\rho\partial_i\delta\Phi-\delta\rho\partial_i\delta\Phi\nonumber\\
&=-\rho\partial_i\Phi-\delta\rho\partial_i\Phi-\rho\partial_i\delta\Phi,\label{eq:mom4}
\end{align}
where we have neglected terms higher-order in the perturbations. 

Collecting the terms given by Equations~\eqref{eq:mom1}\eqref{eq:mom2},\eqref{eq:mom3}, and \eqref{eq:mom4}, the momentum equation becomes
\begin{align}
\left[\partial_t\right.&\left.(\rho v_i)+\nabla_j(\rho v^j v_i)+\partial_iP+\rho\partial_i\Phi\right]\nonumber\\
&+\rho\gamma_{ij}\partial_t\delta v^j+\partial_i\delta P=-\delta\rho\partial_i\Phi-\rho\partial_i\delta\Phi.
\end{align}
The term in square brackets contains only background terms, which satisfy Equation~\eqref{eq:momentum}, and so it vanishes. We are thus left with
\begin{equation}
    \rho\gamma_{ij}\partial_t\delta v^j+\partial_i\delta P=-\delta\rho\partial_i\Phi-\rho\partial_i\delta\Phi.
\end{equation}
Expressing the velocity perturbation in terms of the fluid displacement vector, we have
\begin{equation}
    \rho\gamma_{ij}\partial^2_t\xi^j+\partial_i\delta P=-\delta\rho\partial_i\Phi-\rho\partial_i\delta\Phi.\label{eq:mom_final}
\end{equation}

Finally, linearizing the Poisson equation, Equation~\eqref{eq:grav}, we find
\begin{align}
    \nabla^2\delta\Phi&=4\pi\delta\rho-\nabla^2\Phi+4\pi\rho\nonumber\\
                      &=4\pi\delta\rho,\label{eq:grav_final}
\end{align}
where the last equality follows from the fact that the last two terms in the line above satisfy Equation~\eqref{eq:grav} and vanish.

We can now specify that our perturbation is purely azimuthal, as has been done in past studies \cite{West20, MoRaBu18, ZhAnOc24}, and we can then split Equation~\eqref{eq:mom_final} into separate radial and azimuthal equations. Using the components of the flat 3-metric in spherical coordinates and recalling that the background terms have only a radial dependence, we can rearrange Equations~\eqref{eq:cont_final}, \eqref{eq:mom_final}, and \eqref{eq:grav_final} to arrive at the following linear system:
\begin{align}
\delta\rho+\rho\partial_i\xi^i+\rho\xi^i \partial_i\mathrm{ln}\sqrt{\gamma}+\xi^r\partial_r\rho&=0,\label{eq:lincont}\\
\partial^2_t\xi^r+\frac{1}{\rho}\partial_r\delta P+\frac{\delta\rho}{\rho}\partial_r\Phi+\partial_r\delta\Phi&=0,\label{eq:momr}\\
r^2\partial^2_t\xi^\theta+\frac{1}{\rho}\partial_\theta\delta P+\partial_\theta\delta\Phi&=0,\label{eq:momtheta}\\
\nabla^2\delta\Phi-4\pi\delta\rho&=0.\label{eq:lingrav}
\end{align}
\subsubsection{Assumption of Hydrostatic Equilibrium}
\label{subsub:HEC}
Here, we make the distinction that our simulation data does not include the actual Newtonian potential $\Phi$, but the pseudo-Newtonian effective potential. Since Equations~\eqref{eq:cont} and \eqref{eq:momentum} are satisfied in both the Newtonian and pseudo-Newtonian frameworks, we eventually intend to use the pseudo-Newtonian quantities for pressure and density. However, we know that Equation~\eqref{eq:grav} is not satisfied for the pseudo-Newtonian effective potential of the simulation. For this reason, we will remove direct dependencies on the purely Newtonian potential $\Phi$ by using the hydrostatic condition of Equation~\eqref{eq:HEC} to replace it in terms of the density $\rho$ and pressure $P$. While Equation~\eqref{eq:HEC} is not satisfied by the simulation data either, inside the PNS the condition is approximately satisfied. 

\begin{figure}[H]
    \centering
    \includegraphics[width=8.5cm]{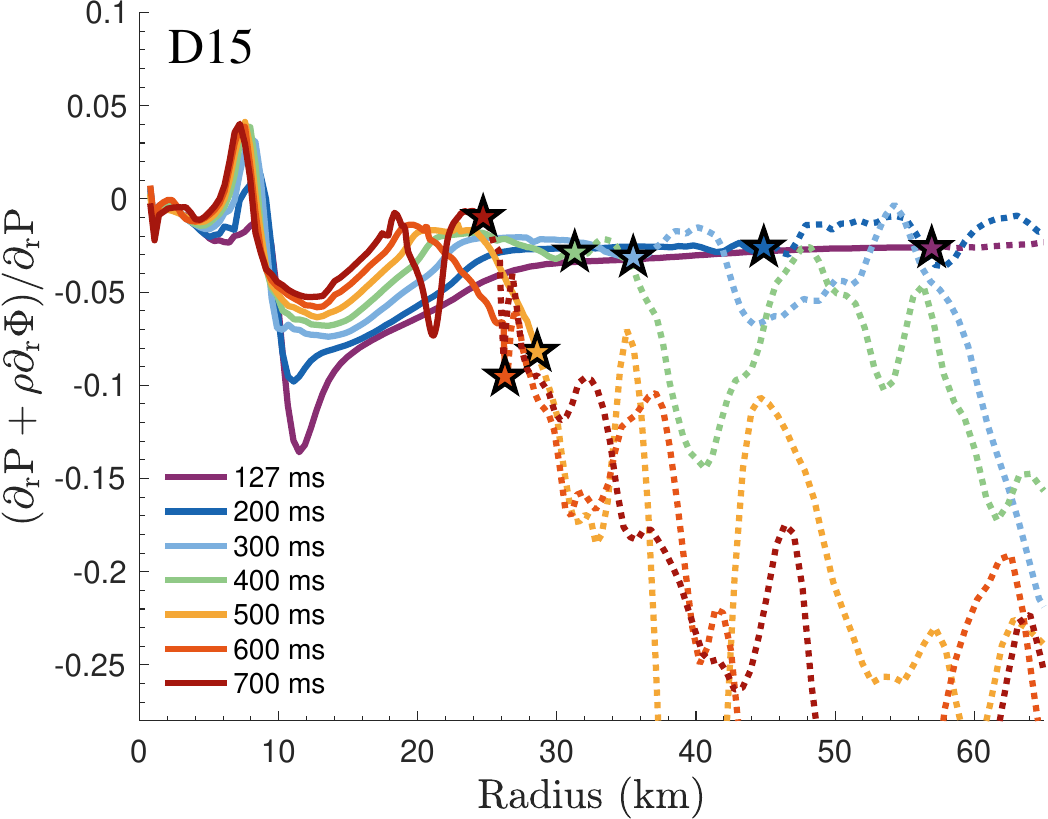}
    \includegraphics[width=8.5cm]{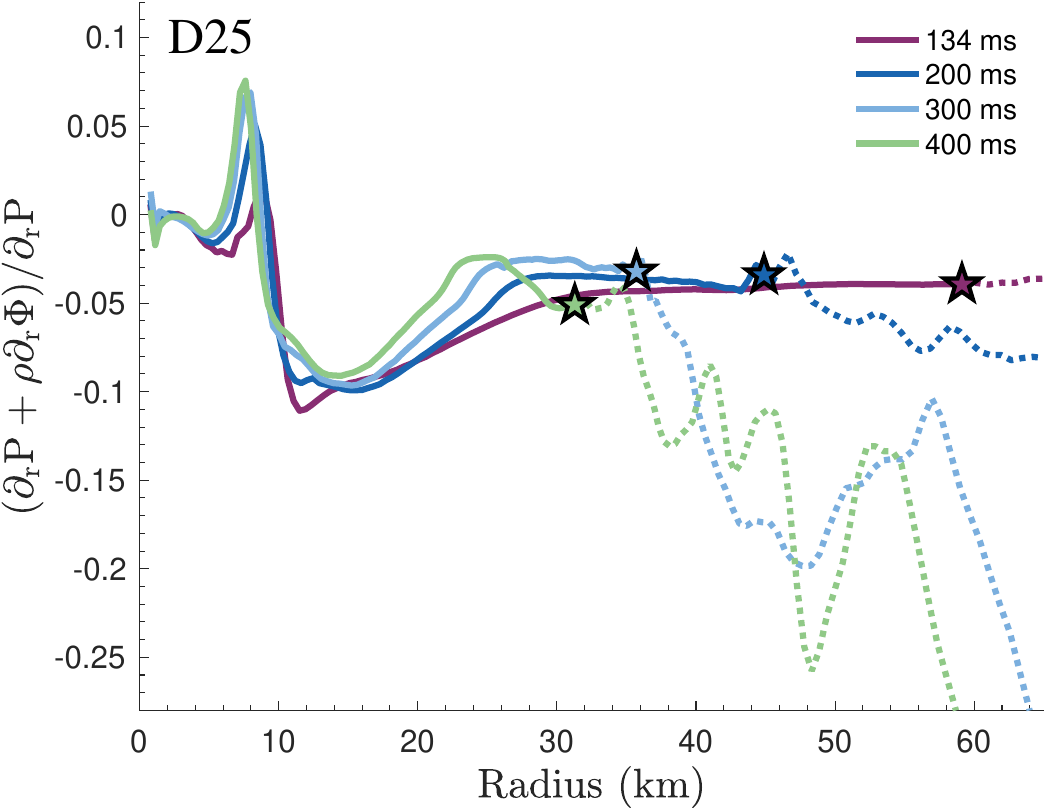}
    \caption{Plots of the deviation from hydrostatic equilibrium, where $\partial_rP=-\rho\partial_r\Phi$, for both D15 and D25. The deviations are plotted at different times differentiated by color. The star symbol indicates the surface of the PNS, defined as the $10^{11}$ g cm$^{-3}$ density contour. The deviation from hydrostatic equilibrium outside the PNS surface is shown by dotted lines.}
    \label{fig:HEC_early}
\end{figure}

Consistent with considerations of the energy-averaged neutrino mean free paths in our D15 and D25 models, we define the PNS surface in this analysis as the $10^{11}$ g cm$^{-3}$ density contour, with its associated radius as $R_{\rm PNS}$. Figure \ref{fig:HEC_early} shows how well the hydrostatic condition is satisfied within the PNS at different times, with the earliest time plotted determined by the start time of the HFF, as defined by \citet{MuCaMe24}.

We see that outside the convective region of D15 at the earliest times, the hydrostatic condition is satisfied to within $\sim$10\%. For D15, the top panel of Figure \ref{fig:HEC_early} shows that, beyond the surface of the PNS, large deviations from hydrostatic equilibrium are observed, particularly at late times. Thus, within the PNS, to a good approximation we can assume hydrostatic equilibrium and rewrite Equation~\eqref{eq:momr} as 
\begin{equation}
    \partial^2_t\xi^r+\frac{1}{\rho}\partial_r\delta P-\frac{\delta\rho}{\rho^2}\partial_rP+\partial_r\delta\Phi=0.\label{eq:momr_final}
\end{equation}

\subsubsection{Linearized System Solutions}
\label{subsub:Solutions}
The linearized system given by Equations~\eqref{eq:lincont}--\eqref{eq:lingrav} can be solved using separation of variables by assuming Anzätze for the variables in the linear system:
\begin{align}
\delta u&=\delta\hat{u}(r)Y_\ell e^{-i\sigma t},\\
\xi^r&=\eta_r(r)Y_\ell e^{-i\sigma t},\\
\xi^\theta&=\frac{\eta_\perp(r)}{r^2}\partial_\theta Y_\ell e^{-i\sigma t},
\end{align}
with $u$ representing either $\rho$, $P$, or $\Phi$, $\sigma$ as the angular frequency, and $Y_\ell $ being the axisymmetric spherical harmonics. In solving for the Eularian displacement vector, we introduce the radial component of the amplitude of the displacement as $\eta_r$ and the azimuthal component of the amplitude of the displacement as $\eta_\perp$. The angular frequency of the perturbation, $\sigma$, is related to the linear frequency, $f$, by  $\sigma=2\pi f$, with $f$ given in units of Hz.

We can eliminate the dependence on $\delta P$ by inserting these Anzäzte into Equation~\eqref{eq:momtheta}, to find
\begin{align}
    0&=r^2\partial_t^2\left(\frac{\eta_\perp}{r^2}\partial_\theta Y_\ell e^{-i\sigma t}\right)
    +\frac{1}{\rho}\partial_\theta\left(\delta\hat{\rho}Y_\ell e^{-i\sigma t}\right)\nonumber\\&\quad+\partial_\theta\left(\delta\hat{\Phi}Y_\ell e^{-i\sigma t}\right)\nonumber\\
    &=-\sigma^2\eta_\perp e^{-i\sigma t}\partial_\theta Y_\ell
    +\frac{\delta \hat{P}}{\rho}e^{-i\sigma t}\partial_\theta Y_\ell \nonumber\\&\quad+\delta\hat{\Phi}e^{-i\sigma t}\partial_\theta Y_\ell.
\end{align}
For nontrivial solutions, we must have the coefficients of $e^{-i\sigma t}\partial_\theta Y_\ell$ satisfy
\begin{align}
    -\sigma^2\eta_\perp+\frac{1}{\rho}\delta\hat{P}+\delta\hat{\Phi}=0.
\end{align}
Rearranging this equation, we thus find $\delta\hat{P}$ is determined by
\begin{equation}
    \delta\hat{P}=\rho\left(\sigma^2\eta_\perp-\delta\hat{\Phi}\right).\label{eq:P_dep}
\end{equation}

As in all other modal analyses of the PNS, we assume the perturbations are adiabatic. Then
\begin{equation}
    \frac{\Delta P}{\Delta \rho}=c_s^2,
\end{equation}
where $\Delta$ designates a Lagrangian perturbation and $c_s^2$ is the square of the sound speed. The sound speed is related to the density and  pressure by
\begin{equation}
    c_s^2=\frac{\Gamma_1P}{\rho},
\end{equation} 
with $\Gamma_1$ being the local adiabatic index. Lagrangian perturbations can be related to Eulerian perturbations through the Eulerian displacement vector:
\begin{equation}
    \Delta u= \delta u+\xi^i\nabla_iu.
\end{equation}
This leads to an additional elimination of the $\delta \hat{\rho}$ dependence. The adiabatic condition becomes
\begin{align}
    \frac{\delta P+\xi^i\partial_iP}{\delta\rho+\xi^i\partial_i\rho}&=c_s^2,
\end{align}
which, due to the spherically-averaged background, reduces to
\begin{align}
    \delta\rho=\frac{1}{c_s^2}\delta P+\frac{1}{c_s^2}\xi^r\partial_rP-\xi^r\partial_r\rho.\label{eq:dp1}
\end{align}
Plugging in our Anzätze for $\delta\rho$, $\delta P$, and $\xi^r$, Equation~\eqref{eq:dp1} becomes
\begin{align}
    \delta\hat{\rho}Y_\ell e^{-i\sigma t}&=\frac{1}{c_s^2}\delta\hat{P}Y_\ell e^{-i\sigma t}\nonumber\\&\qquad+\frac{1}{c_s^2}\eta_rY_\ell e^{-i\sigma t}\partial_rP-\eta_rY_\ell e^{-i\sigma t}\partial_r\rho,
\end{align}
and, for nontrivial solutions, $\delta\hat{\rho}$ must be determined by
\begin{align}
    \delta\hat{\rho}&=\frac{1}{c_s^2}\delta\hat{P}+\frac{\eta_r}{c_s^2}\partial_rP-\eta_r\partial_r\rho.
\end{align}
Using Equation~\eqref{eq:P_dep}, and the definition of the sound speed, we then have
\begin{align}
    \delta\hat{\rho}&=\frac{1}{c_s^2}\rho\left(\sigma^2\eta_\perp-\delta\hat{\Phi}\right)+\eta_r\frac{\rho}{\Gamma_1}\frac{\partial_rP}{P}-\eta_r\partial_r\rho\nonumber\\
    &=\rho\left(\frac{\sigma^2}{c_s^2}\eta_\perp-\frac{\delta\hat{\Phi}}{c_s^2}+\frac{\eta_r}{\Gamma_1}\frac{\partial_rP}{P}-\eta_r\frac{\partial_r\rho}{\rho}\right)\nonumber\\
    &=\rho\left(\frac{\sigma^2}{c_s^2}\eta_\perp-\frac{\delta\hat{\Phi}}{c_s^2}-\mathcal{B} \eta_r\right)\label{eq:rho_dep},
\end{align}
where we have introduced the Schwarzschild discriminant, defined as 
\begin{equation}
    \mathcal{B}=\partial_r\mathrm{ln}\rho-\frac{1}{\Gamma_1}\partial_r\mathrm{ln}P.
\end{equation}

We can now insert our Anzätze into Equations~\eqref{eq:lincont}, \eqref{eq:momr}, and \eqref{eq:lingrav}, and use Equations~\eqref{eq:P_dep} and \eqref{eq:rho_dep}, to simplify our system of equations. For Equation~\eqref{eq:lincont}, we find
\begin{align}
    0&=\delta\hat{\rho}Y_\ell e^{-i\sigma t}+\rho\partial_r\left(\eta_rY_\ell e^{-i\sigma t}\right)+\rho\partial_\theta\left(\frac{\eta_\perp}{r^2} \partial_\theta Y_\ell e^{-i\sigma t}\right)\nonumber\\&\quad+\rho\eta_rY_\ell e^{-i\sigma t} \partial_r\mathrm{ln}\sqrt{\gamma}+\rho\frac{\eta_\perp}{r^2}\partial_\theta Y_\ell e^{-i\sigma t} \partial_\theta\mathrm{ln}\sqrt{\gamma}\nonumber\\&\quad+\eta_rY_\ell e^{-i\sigma t}\partial_r\rho
\end{align}
and, evaluating $\partial_i\mathrm{ln}\sqrt{\gamma}$, explicitly
\begin{align}
    0&=\delta\hat{\rho}Y_\ell e^{-i\sigma t}+\rho Y_\ell e^{-i\sigma t}\partial_r\eta_r+\rho\frac{\eta_\perp}{r^2} e^{-i\sigma t}\partial^2_\theta Y_\ell \nonumber\\&\quad+\rho\eta_rY_\ell e^{-i\sigma t} \frac{2}{r}+\rho\frac{\eta_\perp}{r^2} e^{-i\sigma t} \mathrm{cot}\theta\partial_\theta Y_\ell \nonumber\\&\quad+\eta_rY_\ell e^{-i\sigma t}\partial_r\rho.
\end{align}
We can use the identity
\begin{equation}
    \partial^2_\theta Y_\ell +\mathrm{cot}\theta\partial_\theta Y_\ell=-\ell(\ell+1)Y_\ell\label{eq:ylid} 
\end{equation}
for $\ell\ne0$ to obtain
\begin{align}
    0&=\delta\hat{\rho}Y_\ell e^{-i\sigma t}+\rho Y_\ell e^{-i\sigma t}\partial_r\eta_r-\rho\frac{\eta_\perp}{r^2} Y_\ell  e^{-i\sigma t}\ell(\ell+1)\nonumber\\ &\qquad+\rho\eta_rY_\ell e^{-i\sigma t}\left(\frac{2}{r}+\frac{\partial_r\rho}{\rho}\right)\nonumber\\
    &=\delta\hat{\rho}+\rho\partial_r\eta_r-\rho\eta_\perp \frac{\ell(\ell+1)}{r^2}+\rho\eta_r\left(\frac{2}{r}+\partial_r\mathrm{ln}\rho\right).
\end{align}
Replacing $\delta\hat{\rho}$ with Equation~\eqref{eq:rho_dep} we see that
\begin{align}
    0&=\rho\left(\frac{\sigma^2}{c_s^2}\eta_\perp-\delta\hat{\Phi}-\mathcal{B} \eta_r\right)\nonumber\\
    &\quad+\rho\partial_r\eta_r-\rho\eta_\perp \frac{\ell(\ell+1)}{r^2}+\rho\eta_r\left(\frac{2}{r}+\partial_r\mathrm{ln}\rho\right)\nonumber\\
    &=\partial_r\eta_r+\left(\frac{2}{r}+\partial_r\mathrm{ln}\rho-\mathcal{B}\right)\eta_r\nonumber\\
    &\quad+\left(\frac{\sigma^2}{c_s^2}-\frac{\ell(\ell+1)}{r^2}\right)\eta_\perp-\frac{1}{c_s^2}\delta\hat{\Phi}.
\end{align}
Using the definition of the Schwarzschild determinant, we then have
\begin{align}
    \partial_r\eta_r&+\left(\frac{2}{r}-\frac{1}{\Gamma_1}\partial_r\mathrm{ln}P\right)\eta_r\nonumber\\
    &+\left(\frac{\sigma^2}{c_s^2}-\frac{\ell(\ell+1)}{r^2}\right)\eta_\perp
-\frac{1}{c_s^2}\delta\hat{\Phi}=0.\label{eq:cont_end}
\end{align}

For Equation~\eqref{eq:momr_final}, we find
\begin{align}
    0&=\partial^2_t\left(\eta_rY_\ell e^{-i\sigma t}\right)+\frac{1}{\rho}\partial_r\left(\delta\hat{P}Y_\ell e^{-i\sigma t}\right)\nonumber\\&\qquad-\frac{\partial_r P}{\rho^2}\delta\hat{\rho}Y_\ell e^{-i\sigma t}+\partial_r\left(\delta\hat{\Phi}Y_\ell e^{-i\sigma t}\right)\nonumber\\
    &=-\sigma^2\eta_rY_\ell e^{-i\sigma t}+\frac{1}{\rho}Y_\ell e^{-i\sigma t}\partial_r\delta\hat{P}\nonumber\\
    &\qquad-\frac{\partial_r P}{\rho^2}\delta\hat{\rho}Y_\ell e^{-i\sigma t}+Y_\ell e^{-i\sigma t}\partial_r\delta\hat{\Phi}\nonumber\\
    &=-\sigma^2\eta_r+\frac{1}{\rho}\partial_r\delta\hat{P}-\frac{\partial_r P}{\rho^2}\delta\hat{\rho}+\partial_r\delta\hat{\Phi}.
\end{align}
Using Equations~\eqref{eq:P_dep} and \eqref{eq:rho_dep}, we can replace $\delta\hat{P}$ and $\delta\hat{\rho}$ so that
\begin{align}
    0&=-\sigma^2\eta_r+\frac{1}{\rho}\partial_r\left[\rho\left(\sigma^2\eta_\perp-\delta\hat{\Phi}\right)\right]\nonumber\\
    &\quad-\frac{\partial_r P}{\rho}\left(\frac{\sigma^2}{c_s^2}\eta_\perp-\frac{\delta\hat{\Phi}}{c_s^2}-\mathcal{B} \eta_r\right)+\partial_r\delta\hat{\Phi}\nonumber\\
    &=-\sigma^2\left(1-\frac{\mathcal{B}\partial_r P}{\sigma^2\rho}\right)\eta_r+\sigma^2\partial_r\eta_\perp\nonumber\\&\quad+\sigma^2\left(\partial_r\mathrm{ln}\rho-\frac{\partial_rP}{\rho c_s^2}\right)\eta_\perp-\left(\partial_r\mathrm{ln}\rho-\frac{\partial_rP}{\rho c_s^2}\right)\delta\hat{\Phi}.\label{eq:step2}
\end{align}
Note that
\begin{align}
    \partial_r\mathrm{ln}\rho-\frac{\partial_rP}{\rho c_s^2}&=\partial_r\mathrm{ln}\rho-\frac{\partial_rP}{\rho}\frac{\rho}{\Gamma_1P}\nonumber\\
    &=\partial_r\mathrm{ln}\rho-\frac{1}{\Gamma_1}\partial_r\mathrm{ln}P\nonumber\\
    &=\mathcal{B}.
\end{align}
We can define the gravitational acceleration as $\partial_r P/\rho=\Tilde{G}$ and further define the Brunt-Väisälä frequency squared as $\Tilde{G}\mathcal{B}=\mathcal{N}^2$. 
Thus, Equation~\eqref{eq:step2} becomes
\begin{align}
    0=-\sigma^2\left(1-\frac{\mathcal{N}^2}{\sigma^2}\right)\eta_r+\sigma^2\partial_r\eta_\perp+\sigma^2\mathcal{B}\eta_\perp-\mathcal{B}\delta\hat{\Phi},
\end{align}
and, rearranging terms, we have
\begin{equation}
    \partial_r\eta_\perp-\left(1-\frac{\mathcal{N}^2}{\sigma^2}\right)\eta_r+\mathcal{B}\eta_\perp-\frac{\mathcal{B}}{\sigma^2}\delta\hat{\Phi}=0.\label{eq:momr_end}
\end{equation}

The same procedure for Equation~\eqref{eq:lingrav} yields
\begin{align}
    0&=\nabla^2\left(\delta\hat{\Phi}Y_\ell e^{-i\sigma t}\right)-4\pi\delta\hat{\rho}Y_\ell  e^{-i\sigma t}\nonumber\\
    &=\nabla^2\left(\delta\hat{\Phi}Y_\ell e^{-i\sigma t}\right)\nonumber\\&\qquad-4\pi\rho\left(\frac{\sigma^2}{c_s^2}\eta_\perp-\frac{\delta\hat{\Phi}}{c_s^2}-\mathcal{B}\eta_r\right)Y_\ell  e^{-i\sigma t}.\label{eq:grav_mid}
\end{align}
The Laplacian can be evaluated as
\begin{align}
    &\nabla^2\left(\delta\hat{\Phi}Y_\ell e^{-i\sigma t}\right)\nonumber\\&=\frac{1}{r^2}Y_\ell e^{-i\sigma t}\partial_rr^2\partial_r\delta\hat{\Phi}+\frac{1}{r^2\mathrm{sin}\theta}\delta\hat{\Phi}e^{-i\sigma t}\partial_\theta\mathrm{sin}\theta\partial_\theta Y_\ell\nonumber\\
    &=\left(\partial_r^2\delta\hat{\Phi}+\frac{2}{r}\partial_r\delta\hat{\Phi}\right)Y_\ell e^{-i\sigma t}\nonumber\\
    &\qquad+\left(\frac{1}{r^2}\partial_\theta^2Y_\ell +\frac{1}{r^2}\mathrm{cot}\theta\partial_\theta Y_\ell \right)\delta\hat{\Phi}e^{-i\sigma t}\nonumber\\
    &=\left(\partial_r^2\delta\hat{\Phi}+\frac{2}{r}\partial_r\delta\hat{\Phi}\right)Y_\ell e^{-i\sigma t}-\ell(\ell+1)\frac{\delta\hat{\Phi}}{r^2}Y_\ell e^{-i\sigma t},\label{eq:lap}
\end{align}
where the last step used the identity in Equation~\ref{eq:ylid}. Substituting Equation~\eqref{eq:lap} for the Laplacian term in Equation~\eqref{eq:grav_mid}, nontrivial solutions require that the coefficients of $Y_\ell  e^{-i\sigma t}$ satisfy
\begin{align}
    \partial^2_r\delta\hat{\Phi}+\frac{2}{r}\partial_r\delta\hat{\Phi}&+\left(\frac{4\pi\rho}{c_s^2}-\frac{\ell(\ell+1)}{r^2}\right)\delta\hat{\Phi}\nonumber\\&+4\pi\rho\mathcal{B}\eta_r-4\pi\rho\frac{\sigma^2}{c_s^2}\eta_\perp=0.
    \label{eq:equation59}
\end{align}
In order to keep our ultimate system of equations to first-order ordinary differential equations, we now define
\begin{equation}
    F\equiv\partial_r\delta\hat{\Phi}.\label{eq:fdef}
\end{equation}
Then, Equation~\eqref{eq:equation59} becomes
\begin{align}
    \partial_rF+\frac{2}{r}F&+\left(\frac{4\pi\rho}{c_s^2}-\frac{\ell(\ell+1)}{r^2}\right)\delta\hat{\Phi}\nonumber\\&+4\pi\rho\mathcal{B}\eta_r-4\pi\rho\frac{\sigma^2}{c_s^2}\eta_\perp=0.\label{eq:grav_end}
\end{align}

We can rearrange Equations~\eqref{eq:cont_end}, \eqref{eq:momr_end}, \eqref{eq:fdef}, and \eqref{eq:grav_end} in order to set up a system of ordinary differential equations in $r$ that can be solved to determine the perturbative quantities. We arrange the system such that
\begin{equation}
    \partial_r\vec{u}=\mathbf{A}(\sigma,r,\rho,P,\Gamma_1)\vec{u},\label{eq:sys}
\end{equation}
where
\begin{equation}
   \vec{u}=(\delta\hat{\Phi},F,\eta_r,\eta_\perp)^T 
\end{equation}
and
\begin{align}
   & \mathbf{A}=\nonumber\\
    &\begin{bmatrix}
        0 & 1 & 0 & 0\\
       \frac{\ell(\ell+1)}{r^2}-\frac{4\pi\rho}{c_s^2} & -\frac{2}{r} & -4\pi\rho\mathcal{B} & 4\pi\rho\frac{\sigma^2}{c_s^2}\\
        \frac{1}{c_s^2} & 0 & \frac{\partial_r\mathrm{ln}P}{\Gamma_1}-\frac{2}{r} & \frac{\ell(\ell+1)}{r^2}-\frac{\sigma^2}{c_s^2}\\
        \frac{\mathcal{B}}{\sigma^2} & 0 & 1-\frac{\mathcal{N}^2}{\sigma^2}& -\mathcal{B}
    \end{bmatrix}.
\end{align}
This system of equations assumes that $\ell\ne0$ and that $\partial_rP/\rho=-\Phi$, as described in the derivation. 
\subsubsection{Boundary Conditions}
\label{subsub:BC}

We can solve this system by integrating along r and enforcing boundary conditions. The inner boundary conditions are found by imposing regularity conditions at the origin \cite{HuRoWr66}, as derived in \citet{West20} for the Newtonian perturbative equations we are using here. They are:
\begin{align}
    \eta_r(r_0)&=A_0r_0^{\ell-1},\label{eq:ib1}\\
    \eta_\perp(r_{0})&=\frac{A_0}{\ell}r_0^\ell,\label{eq:ib2}\\
    \delta\hat{\Phi}(r_{0})&=C_0r_0^\ell,\label{eq:ib3}\\
    \partial_r\delta\hat{\Phi}(r_{0})&=\ell C_0r_0^{\ell-1}, \label{eq:ib4}
\end{align}
where $r_0$ is the inner boundary, $A_0$ is an arbitrary amplitude taken to be $10^{-5}$ for our analysis, as in \citet{West20}, and $C_0$ is found by using a root-finding method to evaluate
\begin{equation}
    \left.\left(\partial_r\delta\hat{\Phi}+\frac{\ell(\ell+1)}{r}\delta\hat{\Phi}\right)\right\vert_{R_{\rm PNS}}=0,\label{eq:obC}
\end{equation}
with $R_{\rm PNS}$ being the radius of the surface of the PNS. Equation~\eqref{eq:obC} is equivalent to Equation (A10) of \citet{West20}, and we use it instead of Equation (A12). As noted by Westernacher-Schneider, the rest mass perturbations $\delta\hat{\rho}$ that escape through $R_{\rm PNS}$ are of small amplitude and mainly leak into different harmonics. Therefore, such perturbations can be ignored. We confirm that using Equation (A12) does not yield appreciable differences relative to the results obtained when using Equation~\eqref{eq:obC}.
For the outer boundary condition, we require the Lagrangian pressure perturbation to vanish:
\begin{equation}
    \left.\Delta P\right\vert_{R_{\rm PNS}}=\left.\left(\rho\sigma^2\eta_\perp-\rho\delta\hat{\Phi}+\eta_r\partial_rP\right)\right\vert_{R_{\rm PNS}}=0,\label{eq:obFree}
\end{equation}
corresponding to a free surface. This boundary condition, or the relativistic counterpart, is also used in \cite{MoRaBu18,SoKuTa19,SoTa20b,MoSuTa23,WoFrMi23,ZhAnOc24}, with varying definitions of the PNS surface. Figure \ref{fig:HEC_early} shows that hydrostatic equilibrium below the PNS surface, \textit{i.e.} the $10^{11}$ g cm$^{-3}$ density contour, is approximately satisfied. We do not implement the vanishing radial perturbation amplitude at the shock as a boundary condition, as done in \cite{ToCePa18,ToCePa19}, given that the region around the shock is far outside of hydrostatic equilibrium.

As noted previously, the \chimera\ code enforces spherical symmetry within the first $\sim$8 km of the origin, well below the innermost radius of region 1 in our analysis. In \citet{MeMaLa23}, to ensure that the imposition of spherical symmetry in this limited volume did not impact our findings, we computed the gravitational wave strain in ``region 0,'' whose inner boundary corresponded to the outer boundary of the spherical region and whose outer boundary coincided with the inner boundary of region 1. Negligible gravitational wave strain was found in region 0. Consistent with this treatment, the inner boundary of our modal analysis is set to be the inner boundary of region 0. Thus, the eigenfunctions $\eta_r$ and $\eta_\perp$ extend below region 1, but not all the way to the origin.

\subsubsection{Numerical Scheme}
\label{subsub:Scheme}
To solve Equations~\eqref{eq:sys}, we choose a set of frequencies $\sigma=2\pi f$ and integrate out to the surface of the PNS, then use a root finding procedure to determine the $\sigma$ that satisfies the boundary conditions. This means that we have a nested root solve to determine $C_0$ for each frequency we consider. The solution procedure for \chimera\ data at a particular time step proceeds as follows:

\begin{enumerate}
    \item Preset some frequency array $\sigma'$ spanning the range in which we expect to find roots that satisfy Equation~\eqref{eq:obFree}. 
    \begin{enumerate}
    \item We choose a frequency range of 200-2000 Hz to capture the HFF, and had a frequency resolution $\sim$3 Hz.
    \item The root finding will not guarantee all roots are found. Therefore, it is important to have frequency resolution within this array fine enough to capture adjacent eigenfrequencies that may be within $\sim$10 Hz of each other.
    \end{enumerate}
    \item Read in spherically-averaged, hydrodynamic and thermodynamic variables from \chimera\ for the entire radial grid.
    \begin{enumerate}
        \item Compute derivatives of $\rho$ and $P$ using fourth-order finite differencing.
        \item Interpolate all background quantities onto a uniformly-spaced grid from $r_0$ to $R_{\rm PNS}$, using cubic splines.
    \end{enumerate}
    \item For each frequency:
    \begin{enumerate}
        \item Preset some array of $C_0'$.
            \begin{enumerate}
                \item Ensure sufficient resolution to capture the root. We find typical values between $\pm10^{-19}$ when working in geometrized units, with $G=c=1$.
            \end{enumerate}
        \item Set inner boundary conditions using Equations~\eqref{eq:ib1} and \eqref{eq:ib2}.
        \item Integrate Equation~\eqref{eq:sys} to $R_{\rm PNS}$.
        \item Using values of $C_0'$ that bound Equation~\eqref{eq:obC}, determine the proper $C_0$ via the bisection method.
        \item Use $C_0$ to set inner boundary conditions using Equations~\eqref{eq:ib3} and \eqref{eq:ib4}.
        \item Integrate Equation~\eqref{eq:sys} to the surface of the PNS.
    \end{enumerate}
    \item Using values of $\sigma'$ that bound the root of Equation~\eqref{eq:obFree}, use the bisection method to determine $\sigma$. 
    \begin{enumerate}
        \item Repeat step 3 for the new bounding frequency found at each step of the bisection method.
    \end{enumerate}
\end{enumerate}
We implemented a solver using both the ordinary differential equation solver library ODEPACK \cite{Hind19} and a direct integration method using the trapezoid rule. We found good agreement between the methods. The ODEPACK solver yielded negligible increases in the accuracy of the solution, while greatly increasing the computation time.

\section{Results}
\label{sec:results}
\subsection{Regional Decomposition of the Strains}
\begin{figure}[H]
    \centering
    \includegraphics[width=8.5cm]{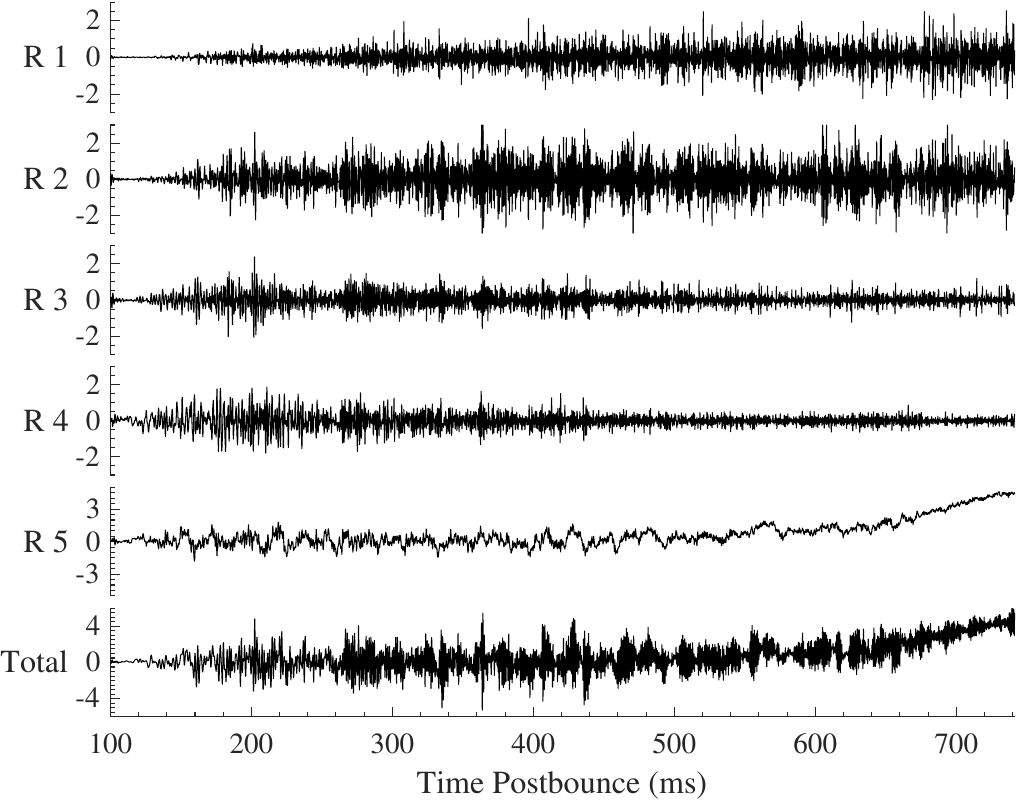}
    \includegraphics[width=8.5cm]{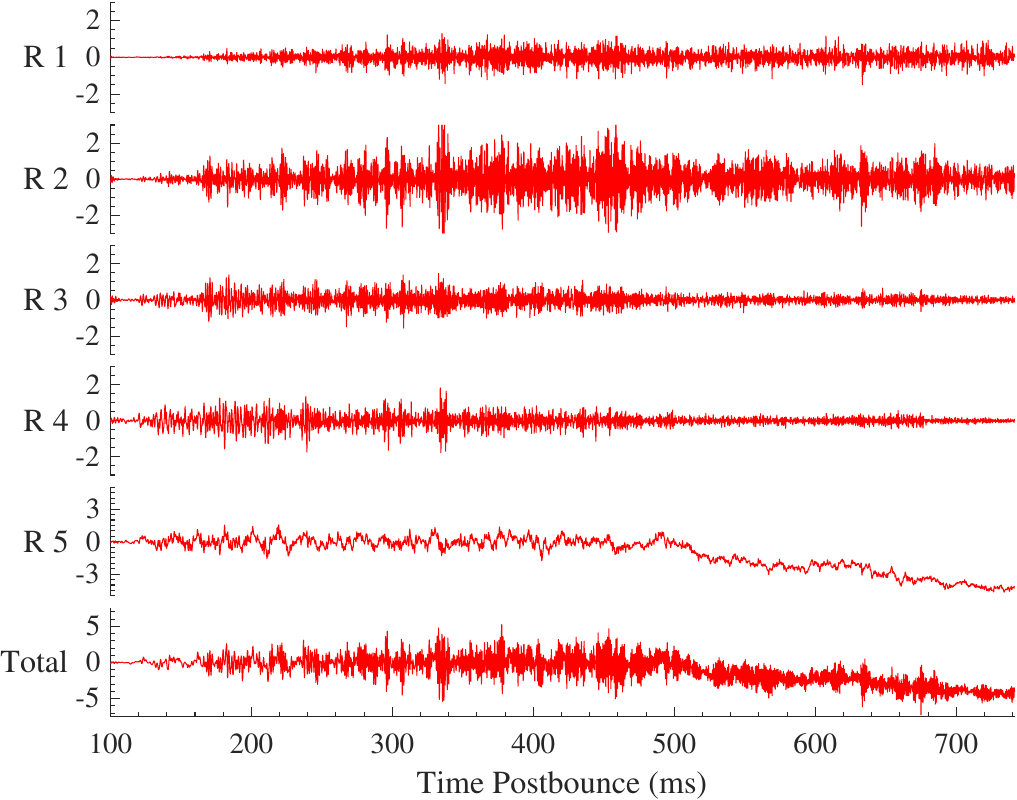}
    \caption{Gravitational wave strains by region for D15 using updated region boundaries as described in the text. The vertical axis shows $Dh_{+,\times}$, with $D=\rm 10$ kpc. Top panel shows the strains for the $h_+$ polarization, bottom panel shows the strains for the $h_\times$ polarization. Note the scales on the vertical axes, as they differ across region and polarization.\\}
    \label{fig:regional_strains_15}
\end{figure}

In Figure \ref{fig:regional_strains_15}, we show the strains for each region, and the total strain, for D15, as determined using the procedure outlined in Sections \ref{sub:GWS} and \ref{sub:Reg_Dec}. In particular, we note the absence of numerical noise in region 5 that was present in Figures 21 and 22 of \citet{MeMaLa23}. From Figure \ref{fig:regional_strains_15}, we can determine the dominant region(s) producing strong gravitational wave strains during each epoch of the core collapse supernova. We include both the $h_+$ and $h_\times$ strains, following the conclusions of \citet{PaVaPa23}. For the first 100 ms after bounce, the D15 and D25 models do not produce significant strains, and those are not shown. 

\begin{figure}
    \centering
    \includegraphics[width=8.5cm]{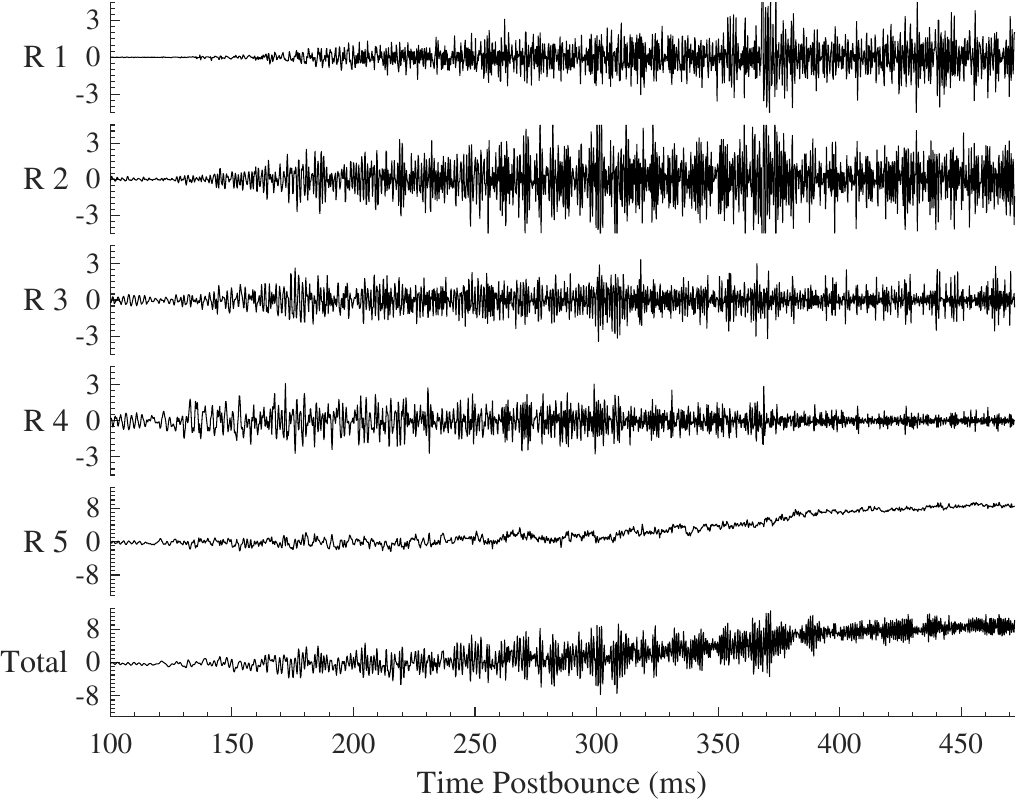}
    \includegraphics[width=8.5cm]{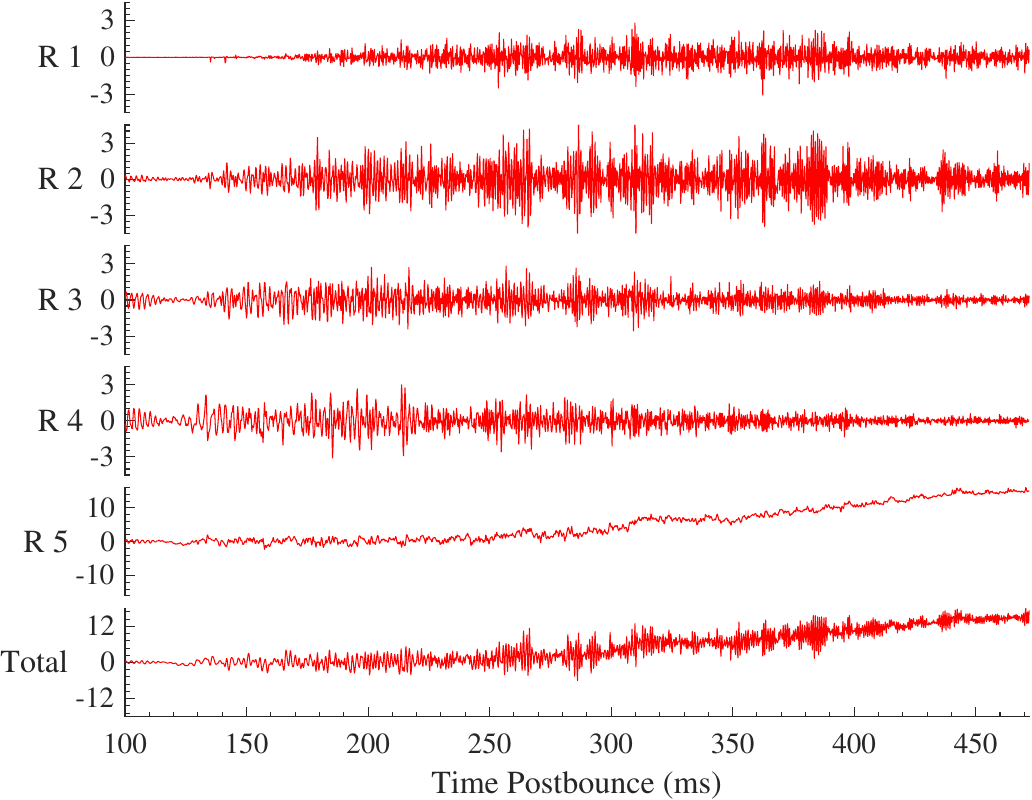}
    \caption{Gravitational wave strains by region for D25 using updated region boundaries as described in the text. The vertical axis shows $Dh_{+,\times}$, with $D=\rm 10$ kpc. Top panel shows the strains for the $h_+$ polarization, bottom panel shows the strains for the $h_\times$ polarization. Note the scales on the vertical axes, as they differ across region and polarization.}
    \label{fig:regional_strains_25}
\end{figure}

For the $h_+$ polarized strains of the D15 model, beginning almost immediately 100 ms after bounce, it is region 4 that initially produces gravitational wave strains followed soon after by region 3, producing strains of a similar amplitude. From 100 to 130 ms, regions 1 and 2 do not show much activity, but 130 ms after bounce they begin to output high-frequency gravitational waves as Ledoux convection sets in \cite{MeMaLa20}. The amplitudes of the strains produced in regions 1 and 2 continue to grow, and at 160 ms postbounce the strains of region 2 are roughly the same amplitude as those of regions 3 and 4. Roughly 240 ms after bounce, region 2 has become the region with the largest-amplitude strains. At 300 ms after bounce, region 1 produces strain amplitudes approximately equal to the amplitudes in regions 3 and 4. As the simulation continues to progress, the amplitudes in regions 3 and 4 decrease as accretion onto the PNS subsides after the explosion begins. Regions 1 and 2 maintain their high-amplitude strains for the rest of the simulation, as the Ledoux convection that drives them continues due to the lepton gradients sustained by neutrino diffusion and emission from the PNS. In Figure \ref{fig:regional_strains_25}, we see that the D25 model shows the same behavior in regional production of gravitational waves with time, although the exact transition times vary between the models. This behavior is clear in both polarizations of the strains. 

From these strains, we see that the general evolution of gravitational wave generation for the D-series models 100 ms after bounce proceeds as: (i) region 4 initially produces small amplitude gravitational wave strains, (ii) region 3 produces strains of similar amplitude a few tens of ms later, (iii) Ledoux convection begins and causes gravitational wave strains to be produced in regions 1 and 2, (iv) within 300 ms after bounce region 1 will have strains of the same magnitude as regions 3 and 4, and region 2 will produce the greatest strains, (v) as accretion onto the surface of the PNS slows, the strains produced in regions 3 and 4 decrease, (vi) sustained Ledoux convection keeps the magnitude of the strains produced in regions 1 and 2 fairly constant for the rest of the simulation. Note that, during phases (iii) and (iv), there is a significant contribution to the total strain across regions 1--4. During these phases, $\sim$200--500 ms for D15 and $\sim$250--380 ms for D25, we see the greatest total strain in both polarizations. Not included in this analysis is the development of standing-accretion-shock-instability (SASI)--produced gravitational waves, which are below $\sim$250 Hz, and the gravitational wave memory, which begins to develop at the time of shock revival, with high amplitude, low-frequency, $\sim$50 Hz and below, gravitational waves. 
\subsection{Fractional Luminosity as a Function of Radius}
This evolution of the gravitational wave production in CCSNe is not only seen in the strains. It is also apparent when examining the gravitational wave luminosity. We can compute the total gravitational wave luminosity as a function of time from \cite{Thor80}
\begin{figure}
    \centering
     \includegraphics[width=8.5cm]{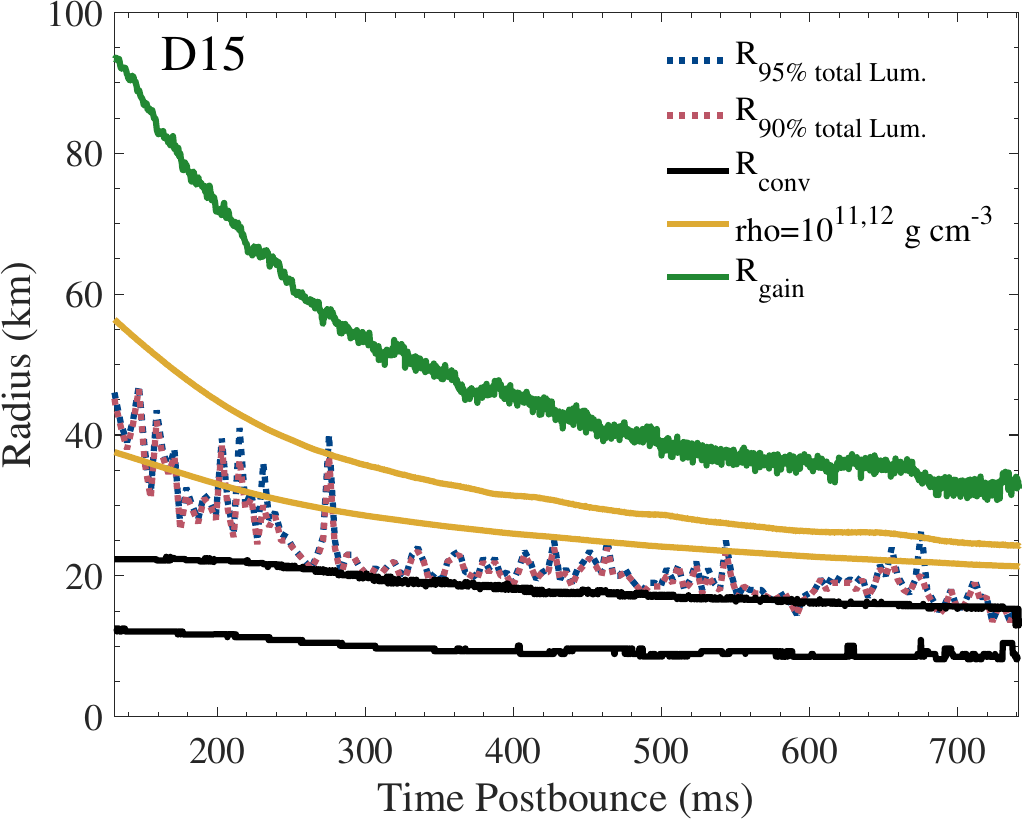}
     \includegraphics[width=8.5cm]{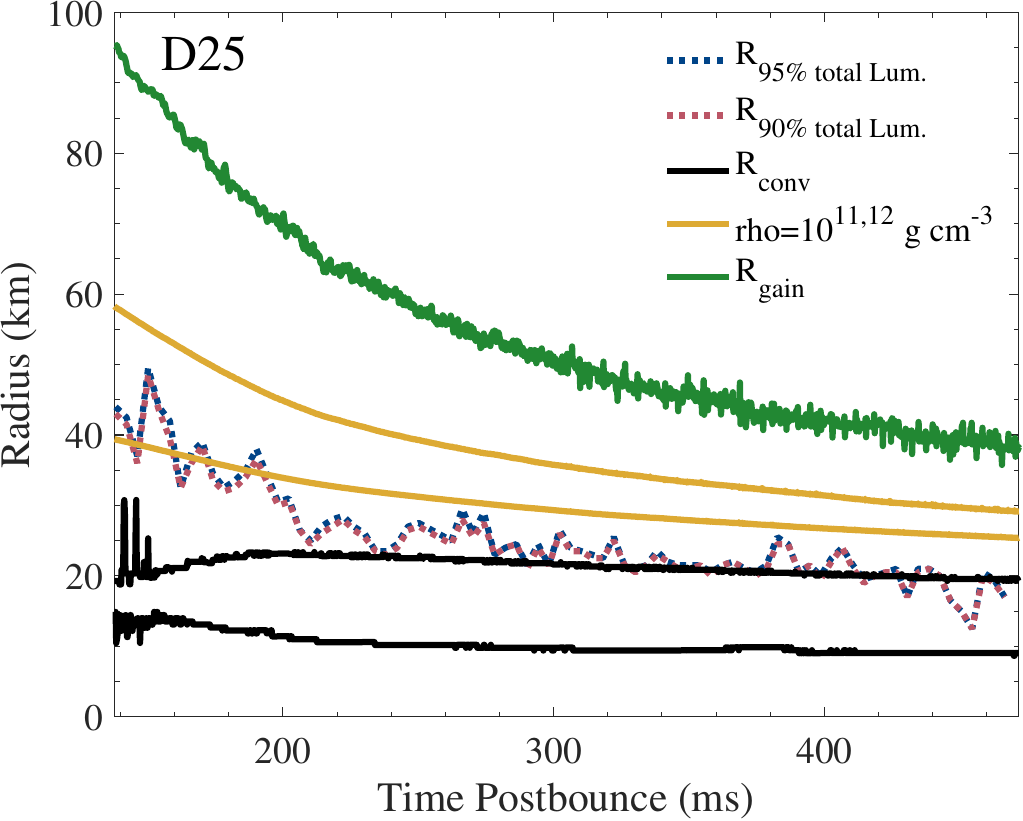}
    \caption{Fractional gravitational wave luminosity as a function of time and radius. The regional boundaries corresponding to the definitions in the text are plotted as a function of time in black, yellow, and green. The red dotted line corresponds to the radius at which 90\% of the gravitational wave luminosity is contained, and the blue dotted line represents the radius at which 95\% of the luminosity is contained. }
    \label{fig:GW_lum}
\end{figure}
\begin{equation}
    \frac{dE}{dt}=\frac{c^3}{G}\frac{1}{32\pi}\sum_{m=-2}^{+2}\left\langle\left|\frac{dA_{2m}}{dt}\right|^2\right\rangle,\label{eq:lum}
\end{equation}
where the $\langle\rangle$ indicate averaging over several wave cycles and
\begin{equation}
    A_{2m}\equiv\frac{dN_{2m}}{dt}=\frac{G}{c^4}\frac{d^2I_{2m}}{dt^2}.
\end{equation}

Due to the nonlinear dependence on $A_{2m}$, it is not possible to decompose the luminosity into regions, as was done for the strains. However, it is possible to compute Equation~\eqref{eq:lum} from the center of the PNS out to a particular radius and compare that to the total gravitational wave luminosity of the entire star. This does not allow for an exact determination of the gravitational wave luminosity within each region of the star, but should allow for an estimate of the fraction of the total luminosity that is emitted within a particular radius, with an obvious correlation then with each region.

Figure \ref{fig:GW_lum} shows the fractional luminosity as a function of radius and time, for each model, for times after the HFF develops. The procedure for determining the start time of the HFF is described in \citet{MuCaMe24}, and is 127 ms and 134 ms after bounce for D15 and D25, respectively. The radii containing 90\% and 95\% of the gravitational wave luminosity are plotted in red and blue, respectively. The fact that these radii lie on top of each other shows that the majority of the gravitational wave luminosity is well bounded by these radii, and that the radii beyond the red and blue curves do not contribute much to the total gravitational wave luminosity. 

In both models, we see that the majority of the gravitational wave luminosity stems from region 3 to the center of the PNS at the onset of the HFF. By 300 ms after bounce for the D15 model and 200 ms after bounce for the D25 model, the majority of the gravitational wave luminosity originates from below region 3. From that time until the end of the simulation, the majority of the gravitational wave luminosity originates from regions 1 and 2.
This is consistent with the evolution of the gravitational wave strains in our regional analysis, discussed earlier.
\begin{figure*}
    \centering
     \includegraphics[width=8.5cm]{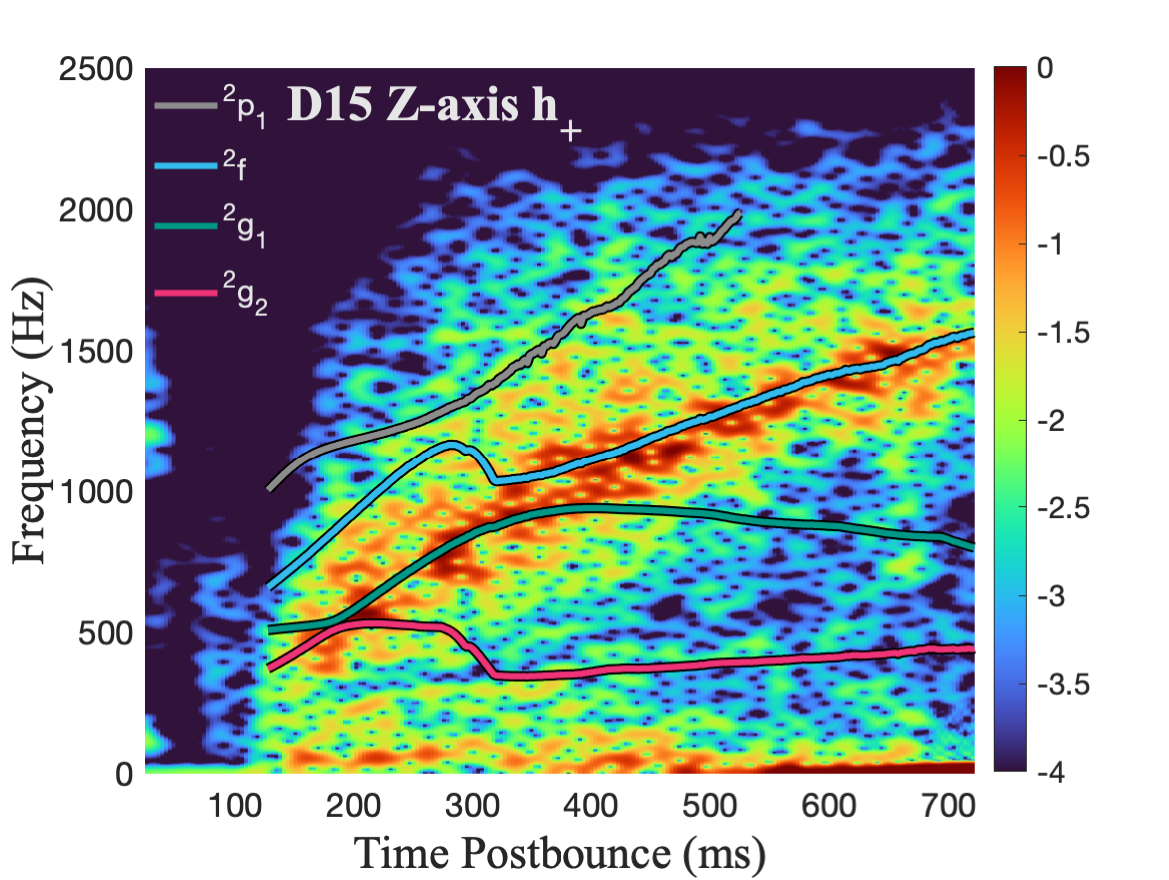}\hfill
     \includegraphics[width=8.5cm]{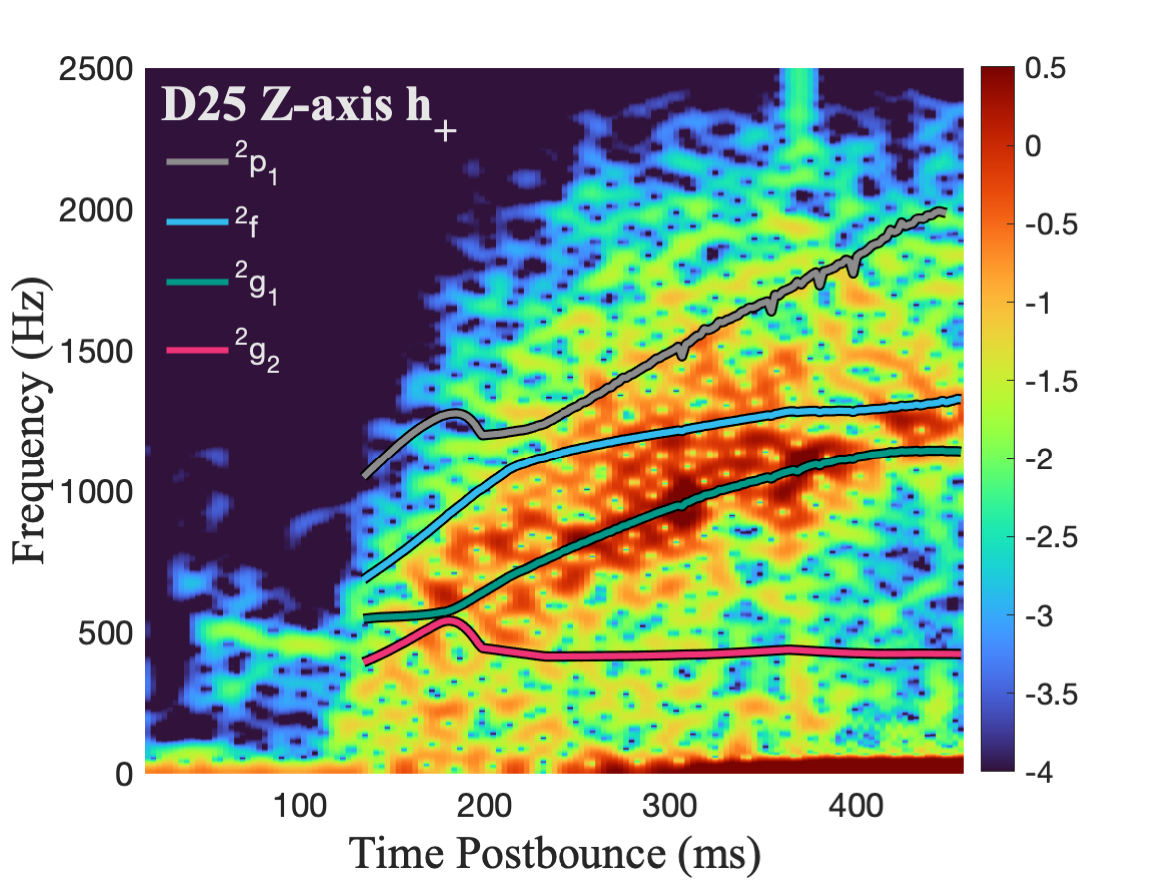}
      \includegraphics[width=8.5cm]{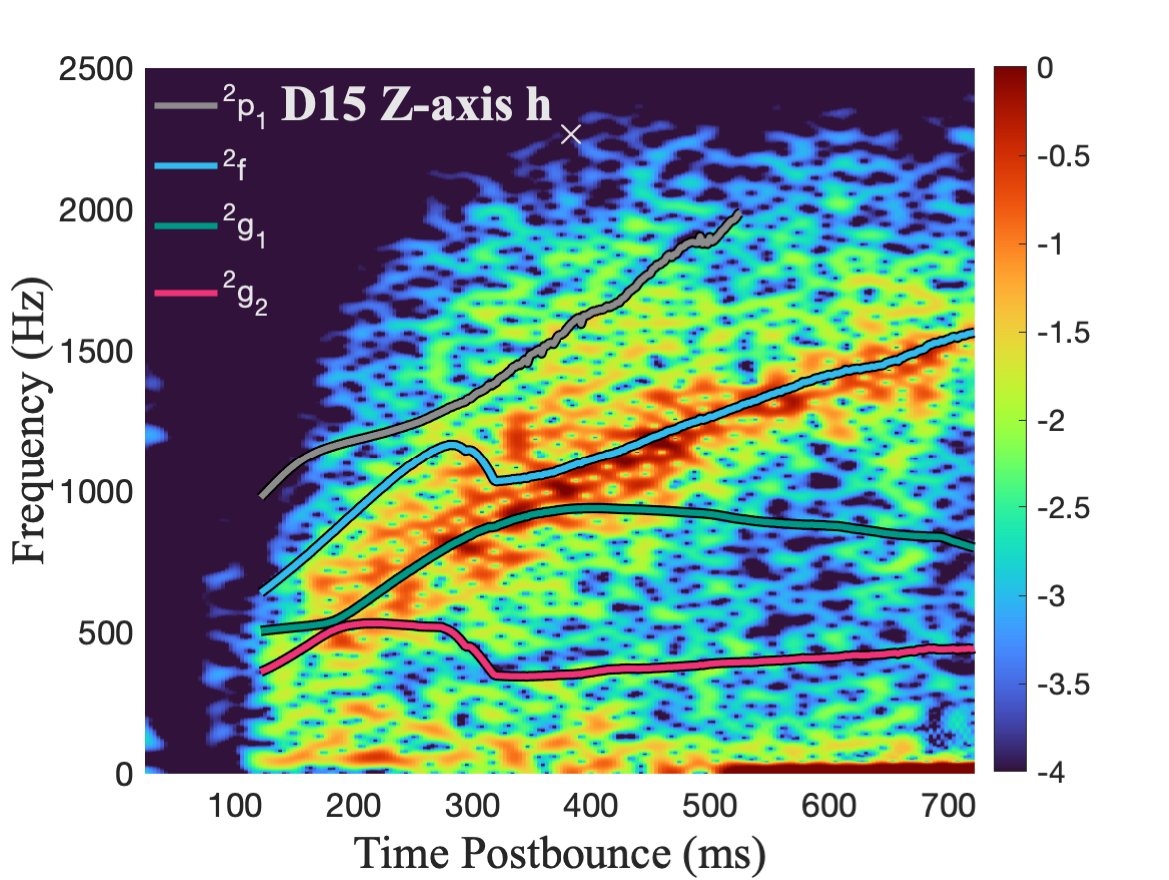}\hfill
     \includegraphics[width=8.5cm]{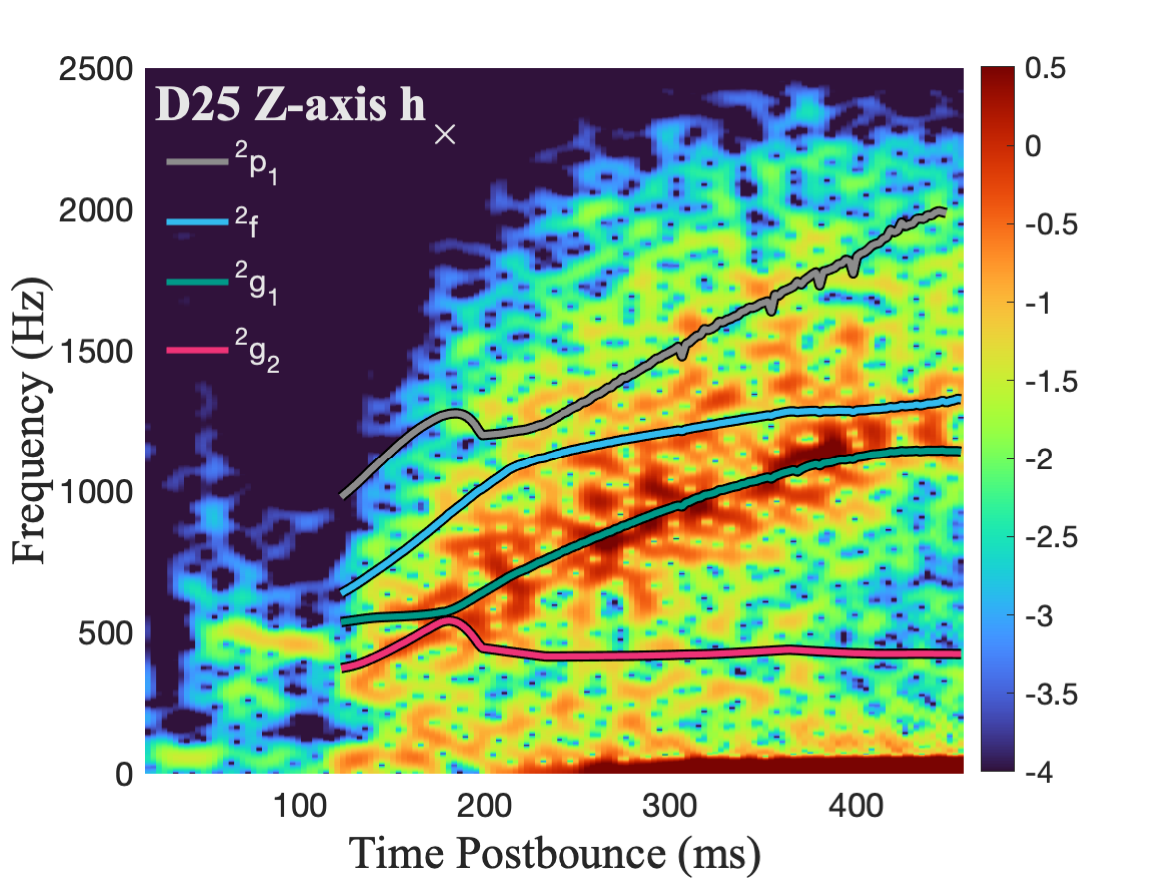}
    \caption{Spectrograms of the gravitational wave emission from models D15 and D25, with the color axis representing the logarithm of the power spectral density $\rm{log}_{10}(P)$. Note the differing color axes between the models, with the higher-mass progenitor resulting in a higher power spectral density. The top row shows the spectrograms of the $h_+$ signals, and the bottom row shows the spectrograms of the $h_\times$ signals. Overlaid on each spectrogram are the eigenfrequencies for each mode that lies close to the HFF, labeled according to the GCN classification as described in the text.}
    \label{fig:Mode_comp}
\end{figure*}
\subsection{Modal Analysis}
The goal of our modal analysis is to find the vibrational modes of the PNS that give rise to its gravitational wave emission. This means that the resulting eigenfunctions for the eigenfrequencies that solve Equation~\eqref{eq:sys} need to be consistent with the qualitative behavior we see in the regional strain and fractional luminosity plots of Figures \ref{fig:regional_strains_15}--\ref{fig:GW_lum}. Additionally, the eigenfrequencies should reproduce the qualitative and quantitative behavior, to be discussed, in the spectrograms in Figure \ref{fig:Mode_comp}. Determining whether a mode reproduces this behavior is nontrivial. To simplify our discussion, we will focus on only the vibrational modes that are close in frequency to the HFF, but we note that the modal analysis produces eigenfrequencies both above and below the HFF.

For our modal analysis, we examine only the quadrupolar, $\ell=2$ non-radial modes, given that they are the largest contributors to the gravitational wave signal. The PNS modes are labeled according to the Generalized Cowling Nomenclature (GCN) as described in \citet{Osaki_75}. The GCN labels each eigenmode by the total number of $p$-nodes, $N_p$, and $g$-nodes, $N_g$, identified in the phase diagram of $(\eta_r,\eta_\perp)$, and each mode is uniquely identified by the quantity
\begin{equation}
    \tilde{n}\equiv N_p-N_g,
\end{equation}
which monotonically increases with mode frequency. In the phase diagram, a nodal point---\textit{i.e.} a point for which $\eta_r(r)=0$---is determined to be a $p$-node if the curve produced by the eigenmode is moving counterclockwise at that point. A $g$-node is a nodal point where the curve of the eigenmode is moving clockwise at that point. From the definition of $\tilde{n}$, all $p$-modes have $\tilde{n}>0$, all $g$-modes have $\tilde{n}<0$, an the fundamental $f$-mode has $\tilde{n}=0$.
The number of nodes for a given eigenmode can change throughout the evolution of the PNS without the label for the mode changing. This is the advantage of the GCN classification. \citet{RoRaCh23} investigated this method of classification and found it consistently tracked eigenmodes for extended periods of time. Indeed, when plotting the amplitude of the radial eigenfunction, $\eta_r$, we confirm that each mode we investigate evolves consistently and is labeled as the same feature across time.

To ensure that no nodal points are missed due to small oscillations, we interpolated the \chimera\ data onto a grid of sufficient resolution to construct the eigenfunctions. For our study, we interpolated to a grid of 8192 radial cells, corresponding to a cell width of $\mathcal{O}$(10 m). For D15, we also determined the number of nodes for a grid of 16,382 cells and found no difference relative to the number obtained with the 8,192 cell grid, as well as no significant changes in the eignefunctions and eigenfrequencies. 

The spectrograms for the D-series models presented in \citet{MeMaLa23} have been updated using the methods outlined in Section II.C of \citet{MuCaMe24}. The D25 model uses a window size of 30 ms, while the longer signal of D15 allows us to use a 45 ms window. All other aspects of the windowing scheme are the same as in \citet{MuCaMe24}, meaning that the effective window length is 3 ms for each model. Figure \ref{fig:Mode_comp} shows the resulting spectrograms overlaid with the eigenfrequencies determined by the procedure outlined in Section \ref{sub:PertAnalysis}, closest to the HFF.

 In Figure \ref{fig:Mode_comp}, we see that, for each model, the portion of the HFF with the highest power-spectral density is tracked throughout its evolution by different modes predicted by our modal analysis. As the HFF begins to develop, it is the ${}^2g_2$-mode that tracks it best. Between 180--200 ms after bounce the ${}^2g_2$-mode stops tracking the HFF and the ${}^2g_1$ mode takes over as the best tracking mode. Around 400 ms after bounce, the ${}^2f$-mode begins to follow the HFF most closely. We thus identify three modes that track the HFF throughout its evolution: ${}^2g_2$, ${}^2g_1$, and ${}^2f$. When investigating the effect of the location of the boundary, we did not see agreement between the HFF and the eigenfrequencies determined, when applying the boundary condition in Equation~\eqref{eq:obFree} at the $10^{10}$ g cm$^{-3}$ density contour.

 \begin{figure}
    \centering
    \includegraphics[width=8.5cm]{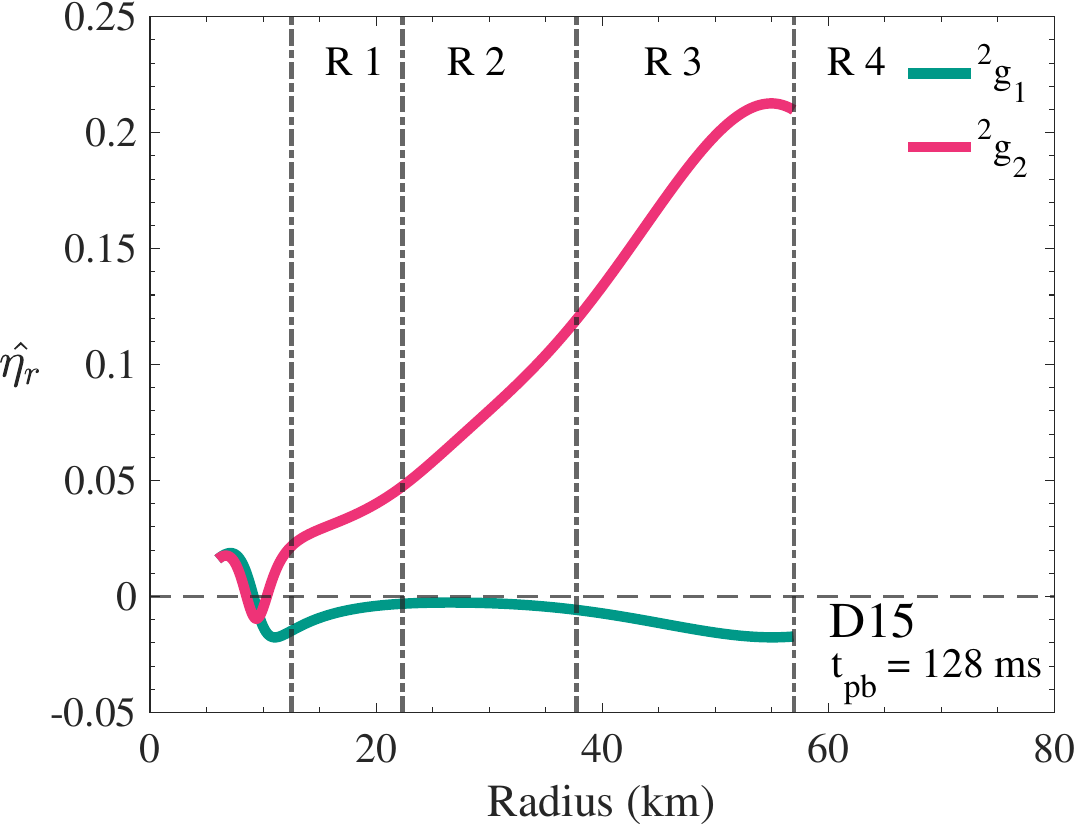}
    \includegraphics[width=8.5cm]{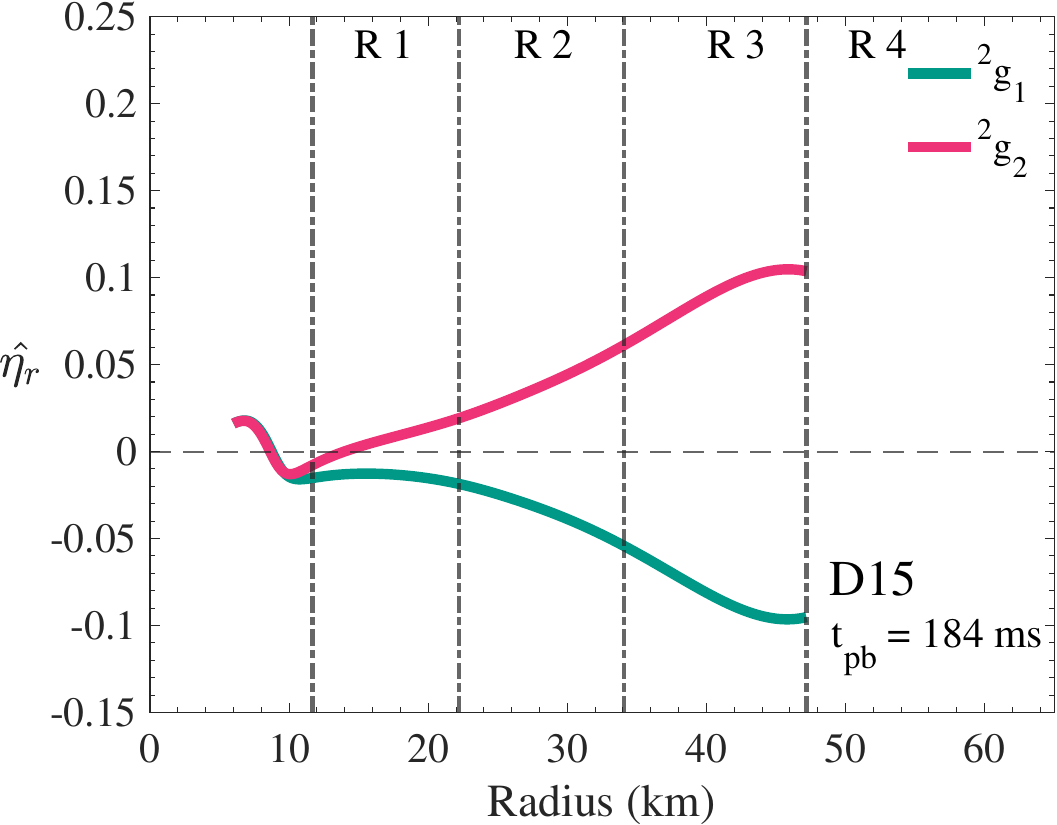}
    \includegraphics[width=8.5cm]{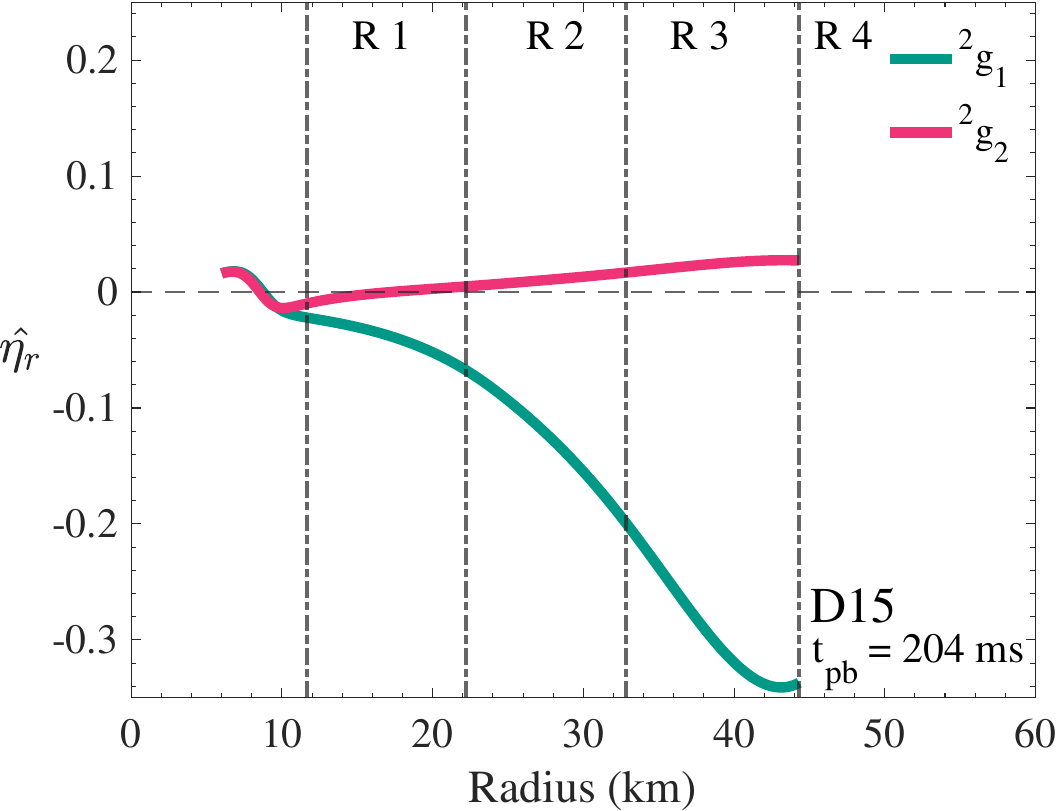}
    \caption{Amplitude of the radial component of the displacement eigenfunction for the ${}^2g_2$- and ${}^2g_1$-modes, around the time of their avoided crossing. The radial boundaries, as defined in the text, are shown as vertical dashed lines. The vertical axis shows the normalized amplitude, $\hat{\eta}_r$, resulting from normalizing with respect to the largest magnitude amplitude across all modes across all times. The approximate avoided crossing time is 184 ms postbounce.} 
    \label{fig:g12_avoid}
 \end{figure}

\begin{figure}
    \centering
    \includegraphics[width=8.5cm]{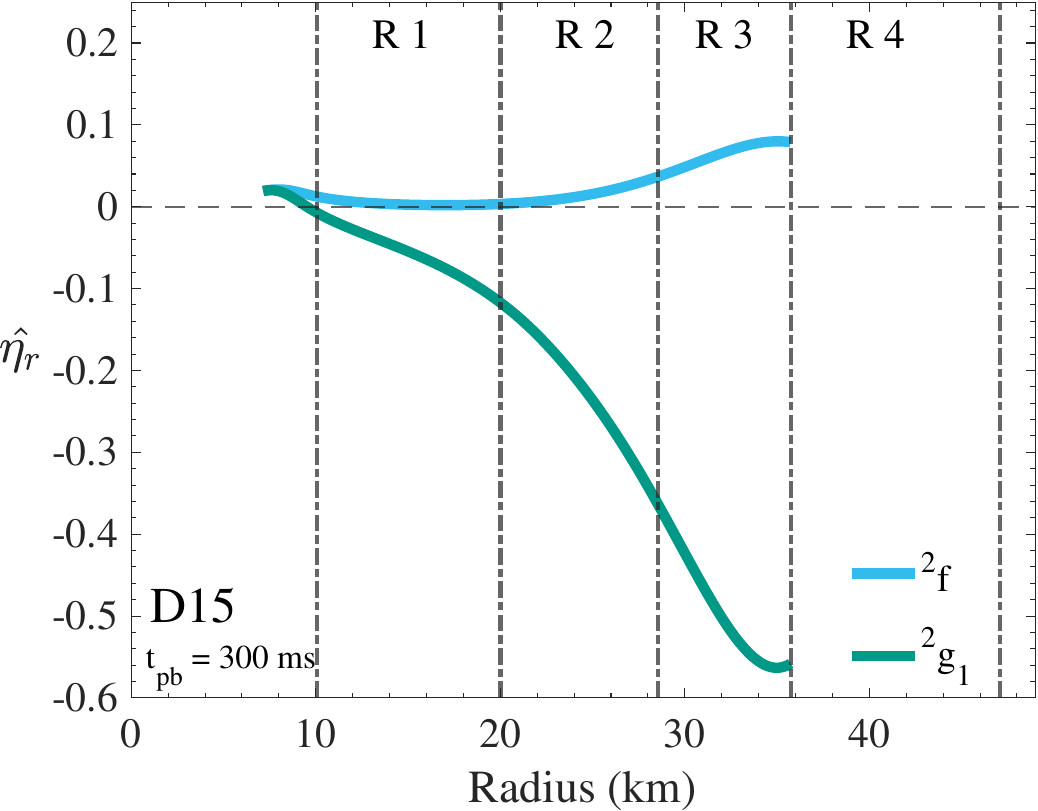}
    \includegraphics[width=8.5cm]{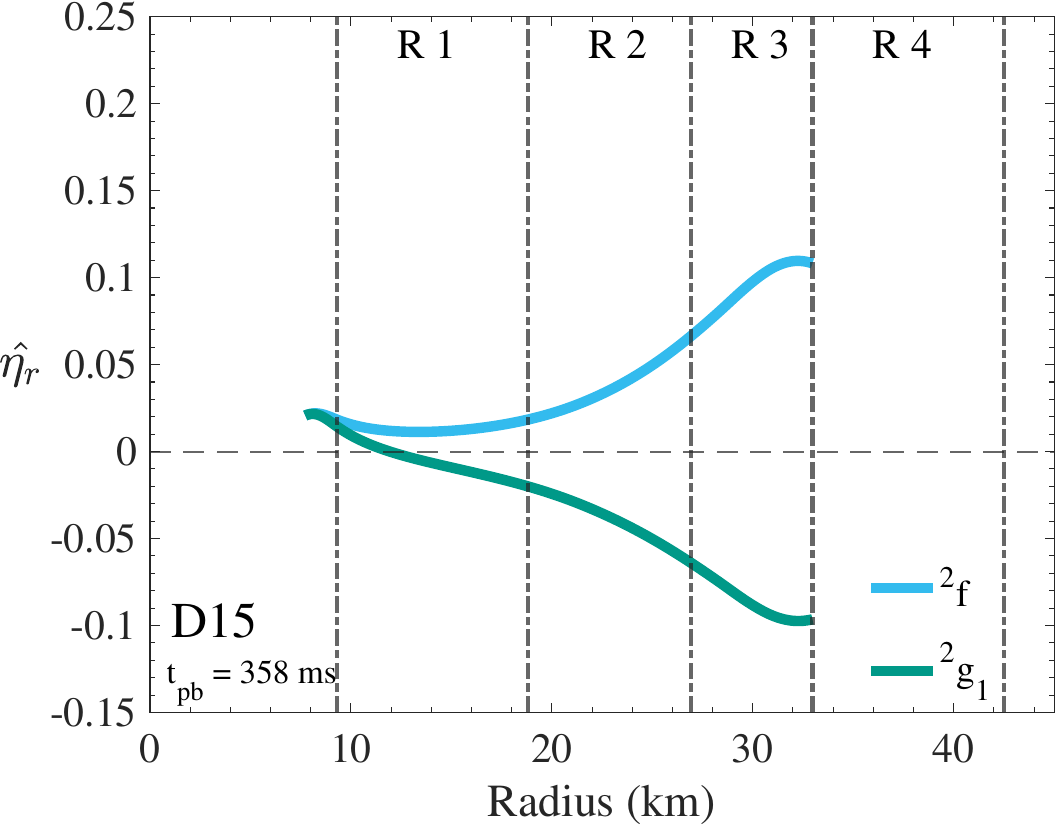}
    \includegraphics[width=8.5cm]{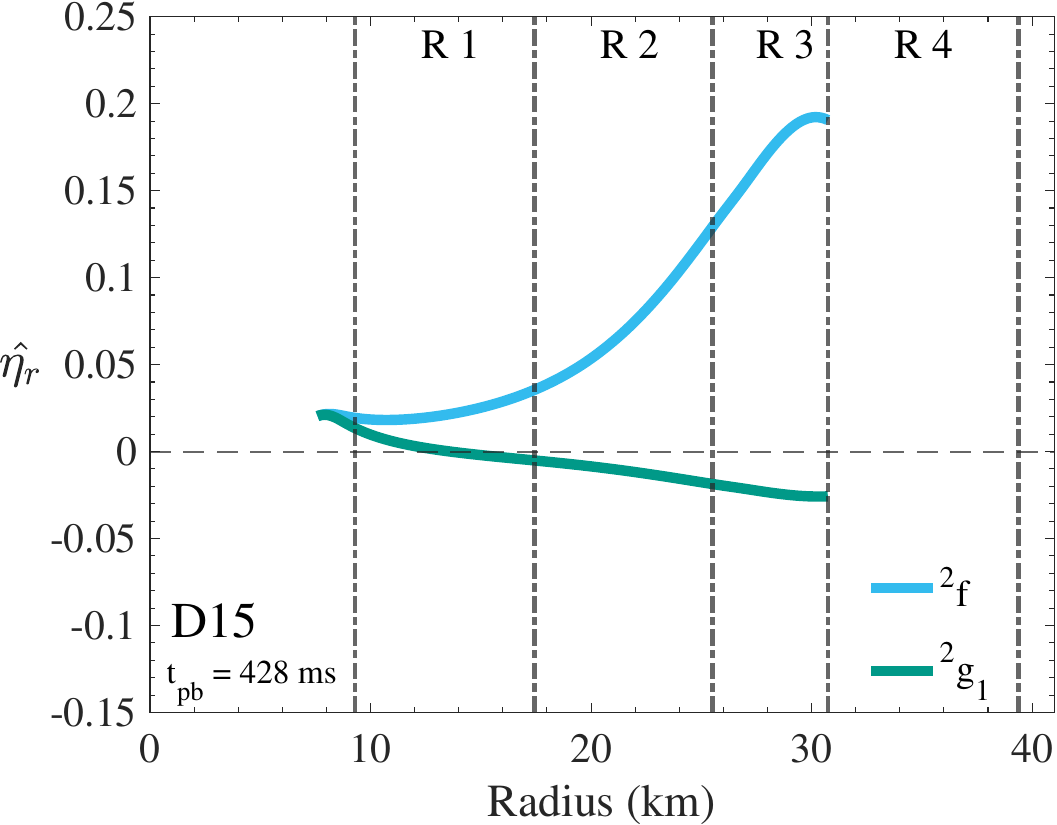}
    \caption{Amplitude of the radial component of the displacement eigenfunction for the ${}^2g_1$- and ${}^2f$-modes, around the time of their avoided crossing. The radial boundaries, as defined in the text, are shown as vertical dashed lines. The vertical axis shows the normalized amplitude, $\hat{\eta}_r$, resulting from normalizing with respect to the largest magnitude amplitude across all modes across all times. The approximate avoided crossing time is 358 ms postbounce.} 
    \label{fig:gf_avoid}
 \end{figure}

 Interestingly, we see that the transition from one mode that tracks the HFF to another occurs at an avoided crossing. In the case of D15, we examine these avoided crossings by plotting the amplitude of the radial component of the perturbation in Figures \ref{fig:g12_avoid} and \ref{fig:gf_avoid}. We see that, before the first avoided crossing shown in Figure \ref{fig:g12_avoid}, the maximum amplitude of the ${}^2g_2$-mode is significantly larger than the maximum magnitude of the ${}^2g_1$-mode amplitude. At the approximate time of the avoided crossing, 184 ms after bounce, the magnitudes of the maximum amplitudes are approximately equal, as the maximum amplitude has decreased for the ${}^2g_2$-mode and increased for the ${}^2g_1$-mode. This trend continues until the ${}^2g_1$-mode has the larger maximum amplitude. This same pattern is seen in the second avoided crossing shown in Figure \ref{fig:gf_avoid}, as well. Here, the maximum magnitude of the amplitude of the ${}^2g_1$ decreases as the maximum amplitude of the ${}^2f$-mode increases. Note that this avoided crossing lasts a much longer time than in the first case. Avoided crossings indicate times at which the character of the modes tracking the HFF involved swap \cite{Asteroseismology_2010}. Thus, these avoided crossings indicate that the character of a mode transitions either to or away from the underlying character of the HFF.

In order to determine if these modes match the evolution of gravitational wave emission as seen in the simulation data of Figures \ref{fig:regional_strains_15}--\ref{fig:Mode_comp}, we investigate the total kinetic energy within a mode, $E_{GW}$, and the power emitted as gravitational waves, $P_{GW}$. To compute the total kinetic energy, we first calculate the kinetic energy density of a mode with frequency $\sigma$ as \cite{ToCePa18,MoRaBu18}
\begin{equation}
    \mathcal{E}(r)=\frac{\sigma^2}{8\pi}\rho\left[\eta_r^2+\ell(\ell+1)\frac{\eta_\perp^2}{r^2}\right].
\end{equation}
The total kinetic energy contained within a mode in the Newtonian limit is then given by
\begin{equation}
    E_{GW}=4\pi\int_0^{R_{\rm PNS}}r^2\mathcal{E}dr.
\end{equation}
The total power radiated from each mode as gravitational waves, in the Newtonian limit, is \cite{Thor69}
\begin{equation}
    P_{GW}=\frac{1}{8\pi}\frac{(\ell+1)(\ell+2)}{(\ell-1)l}\left[\frac{4\pi\sigma^{\ell+1}}{(2\ell+1)!!}\int_0^{R_{\rm PNS}}\delta\hat{\rho}r^{l+2}dr\right]^2,\label{eq:power}
\end{equation}
where $\delta\hat{\rho}$ is defined in Equation \eqref{eq:rho_dep}. Just as in the case of gravitational wave luminosity, the nonlinearity with respect to $\delta\hat{\rho}$ makes clear we cannot use Equation~\eqref{eq:power} to calculate the power output regionally. However, we can introduce the gravitational wave production efficiency defined by \citet{ToCePa18} as
\begin{equation}
    \eta_{GW}=\frac{P_{GW}}{E_{GW}f},
\end{equation}
and use the regional energy contained within a mode as
\begin{align}
    E_{GW}&=4\pi\left[\int_{R1}r^2\mathcal{E}dr+\int_{R2}r^2\mathcal{E}dr+\int_{R3}r^2\mathcal{E}dr\right]\nonumber\\
    &=E_1+E_2+E_3
\end{align}
to decompose the total power output in gravitational waves as
\begin{align}
    P_{GW}&=\eta_{GW}fE_{GW}\nonumber\\
    &=\eta_{GW}f(E_1+E_2+E_3)\nonumber\\
    &=P_1+P_2+P_3.
\end{align}
\begin{figure}[H]
    \centering
     \includegraphics[width=8.5cm]{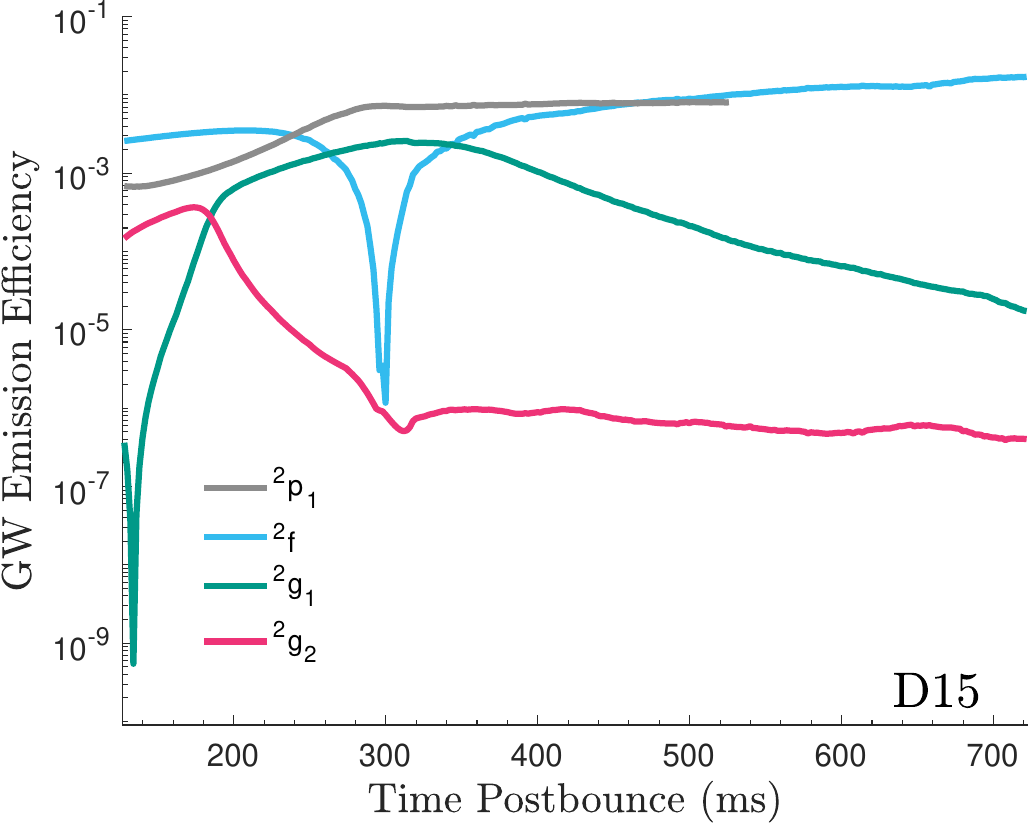}
     \includegraphics[width=8.5cm]{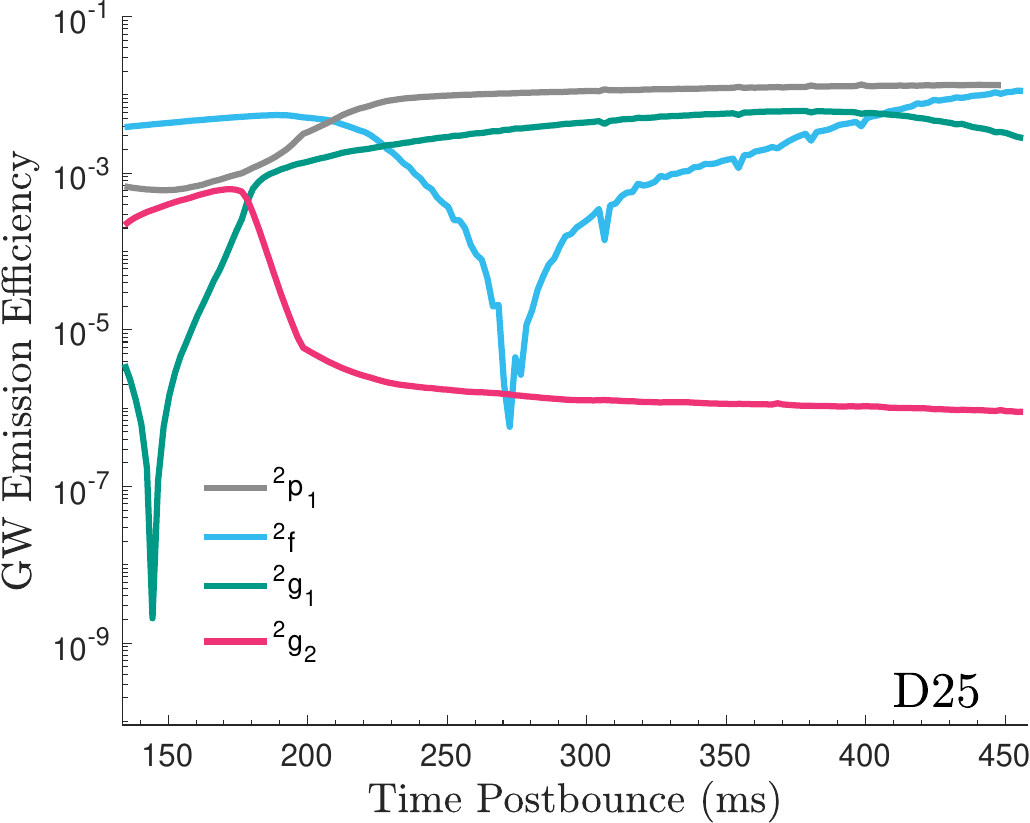}
    \caption{The efficiency of producing gravitational waves as defined in the text for the same modes presented in Figure \ref{fig:Mode_comp} with the same legend. We see that across the two models, after 100 ms, the modes producing gravitational waves most efficiently, and coincide with the HFF in Figure \ref{fig:Mode_comp}, proceed from ${}^2g_2$ to ${}^2g_1$ to ${}^2f$.}
    \label{fig:GW_eff}
\end{figure}

Figure \ref{fig:GW_eff} shows the resulting efficiency of gravitational wave production of each mode for each model. With the exception of the ${}^2f$- and ${}^2p_1$-modes at early times, the times where a mode is most efficient correspond directly to the times where they best fit the HFF, for both models. It is unclear why the ${}^2f$-mode is shown to be so efficient at early times when the spectrograms in Figure \ref{fig:Mode_comp} do not show a significant feature at those frequencies, for either model. However, we do note that the ${}^2f$-mode has two nodal points, one $p$-node and one $g$-node, during this time of overestimated efficiency. As the magnitude of the amplitude of the radial component of the eigenfunction decreases between these two nodal points, the efficiency of the mode begins to decrease as well. By 300 ms after bounce, the ${}^2f$-mode has no nodal points as shown in Figure \ref{fig:gf_avoid}. Additionally, in both models a ${}^2p_1$-mode is observed to have a gravitational wave emission efficiency higher than that of both the ${}^2g_1$- and the ${}^2f$-modes until $\sim$450--460 ms after bounce when the ${}^2f$-mode becomes the highest efficiency mode. We only tracked modes up to 2000 Hz; hence, the artificial termination of the ${}^2p_1$-mode in the plot. We plot the eigenfrequencies of the ${}^2p_1$-mode in Figure \ref{fig:Mode_comp} and see that it does not correspond to any particular feature in the spectrograms. We note that \citet{ToCePa18} see similar behavior, where the ${}^2p_4$-mode is the most efficient emitter of gravitational waves in their model but the ${}^2p_1$ most closely tracks the highest amplitude feature in their spectrogram. At this time we have not investigated what physical mechanism may be responsible for the reduction in efficiency for these modes. As the ${}^2p_1$-mode does not contribute to the HFF, it is not further investigated in this study.

\begin{figure}
    \centering
     \includegraphics[width=8.5cm]{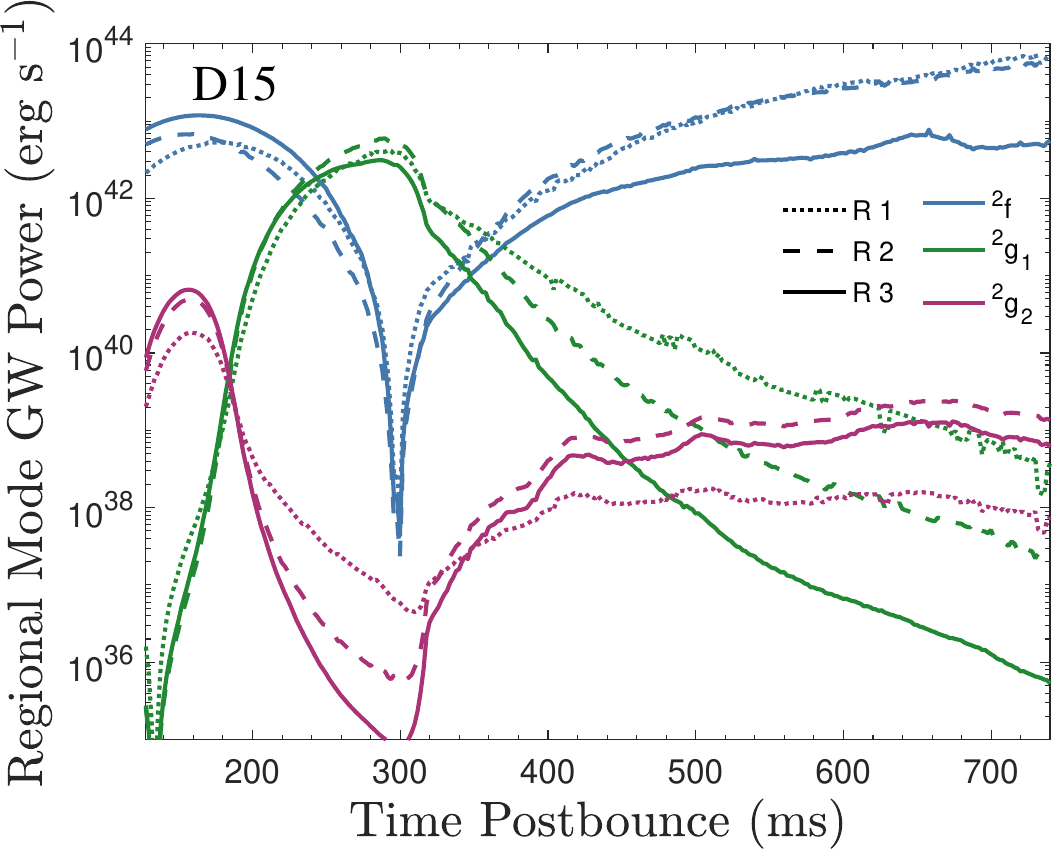}
     \includegraphics[width=8.5cm]{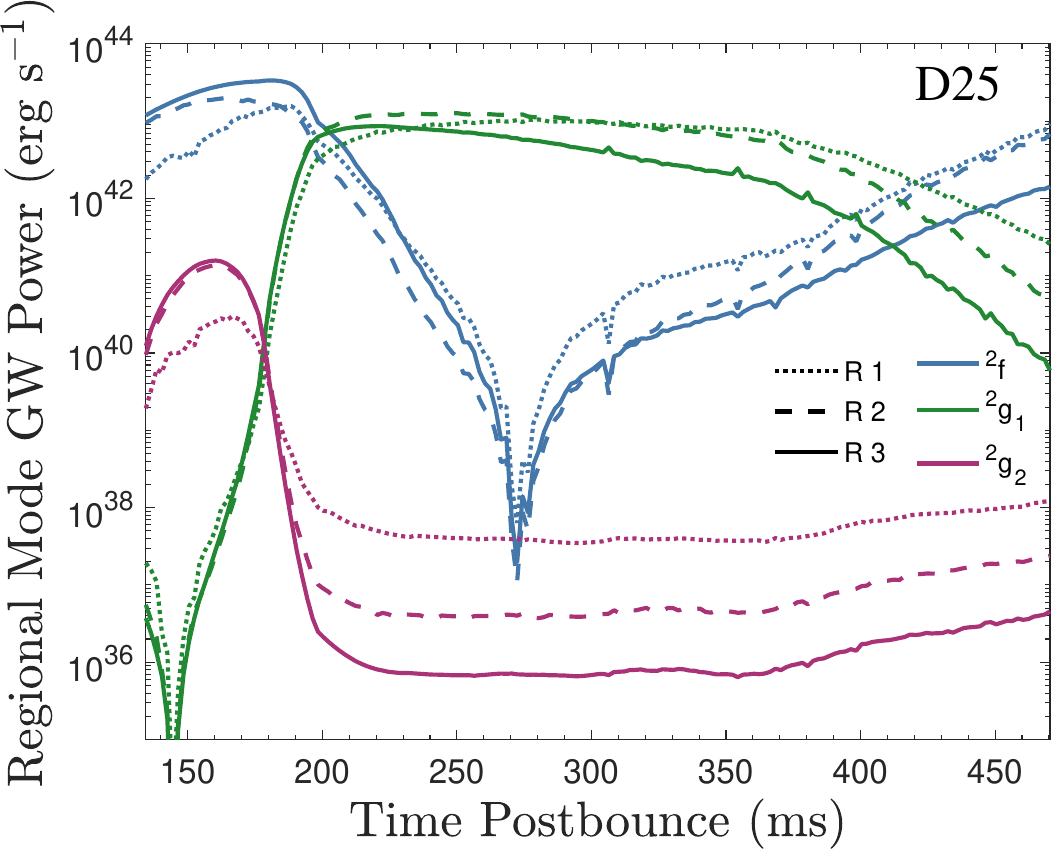}
    \caption{Plots of the approximate power emitted as gravitational waves calculated using Equation~\eqref{eq:power_region}. Results are separated by region for each mode contributing to the HFF. Mode label is denoted by color with ${}^2f$ denoted by blue, ${}^2g_1$ denoted by green, and ${}^2g_2$ denoted by purple. The region in which the power output is approximated is denoted by line type, with region 1 denoted by a dotted line, region 2 denoted by a dashed line, and region 3 denoted by a solid line.}
    \label{fig:Mode_power}
\end{figure}

For the modes that do match the evolution of the HFF, we compute the approximate regional power emitted as gravitational waves for each region $n$ as
\begin{equation}
    P_n=\eta_{GW}fE_n.\label{eq:power_region}
\end{equation} 
Figure \ref{fig:Mode_power} shows the power radiated as gravitational waves for each mode by color, and the regions in which the power is radiated are denoted by line type. Note that the vertical axis is on a log scale, to capture the full range of the power radiated as gravitational waves. Before $\sim$210 ms for the D25 model and before $\sim$240 ms for the D15 model, the power radiated as gravitational waves is concentrated in region 3, denoted by a solid line, regardless of what mode is emitting the most power at that time. Given that the modal analysis uses the surface of the PNS as the outer boundary, we cannot use it to investigate region 4, but from Figure \ref{fig:GW_lum}, we expect the gravitational wave luminosity to be dominated by region 3 even when the strains are slightly larger in region 4. At $\sim$240 ms after bounce for the D15 model, and $\sim$210 ms after bounce for the D25 mode, the dominant power output is from the ${}^2g_1$-mode shown in green, and the majority of the power is emitted in region 2, shown as a dashed line. For the D15 model at $\sim$350 ms after bounce, the ${}^2g_1$-mode begins to radiate power more strongly from region 1, and at $\sim$360 ms after bounce, the ${}^2f$-mode, denoted in blue, becomes the dominant emitter of gravitational waves, with the majority of power coming from region 2. The transition of dominant gravitational wave production from region 2 to region 1 for D25 also occurs at $\sim$330 ms, but when the ${}^2f$-mode becomes dominant at $\sim$420 ms, region 1 remains the dominant region. At the end of the D15 simulation, the power output of regions 1 and 2 are roughly equal, with region 1 being slightly larger. From Figures \ref{fig:g12_avoid} and \ref{fig:gf_avoid}, we see that these transition times correspond approximately to the avoided crossing times as well, again highlighting the fact that these transitions are caused caused by a switch in the character of the modes tracking the HFF.

\begin{figure*}
    \centering
     \includegraphics[width=8.5cm]{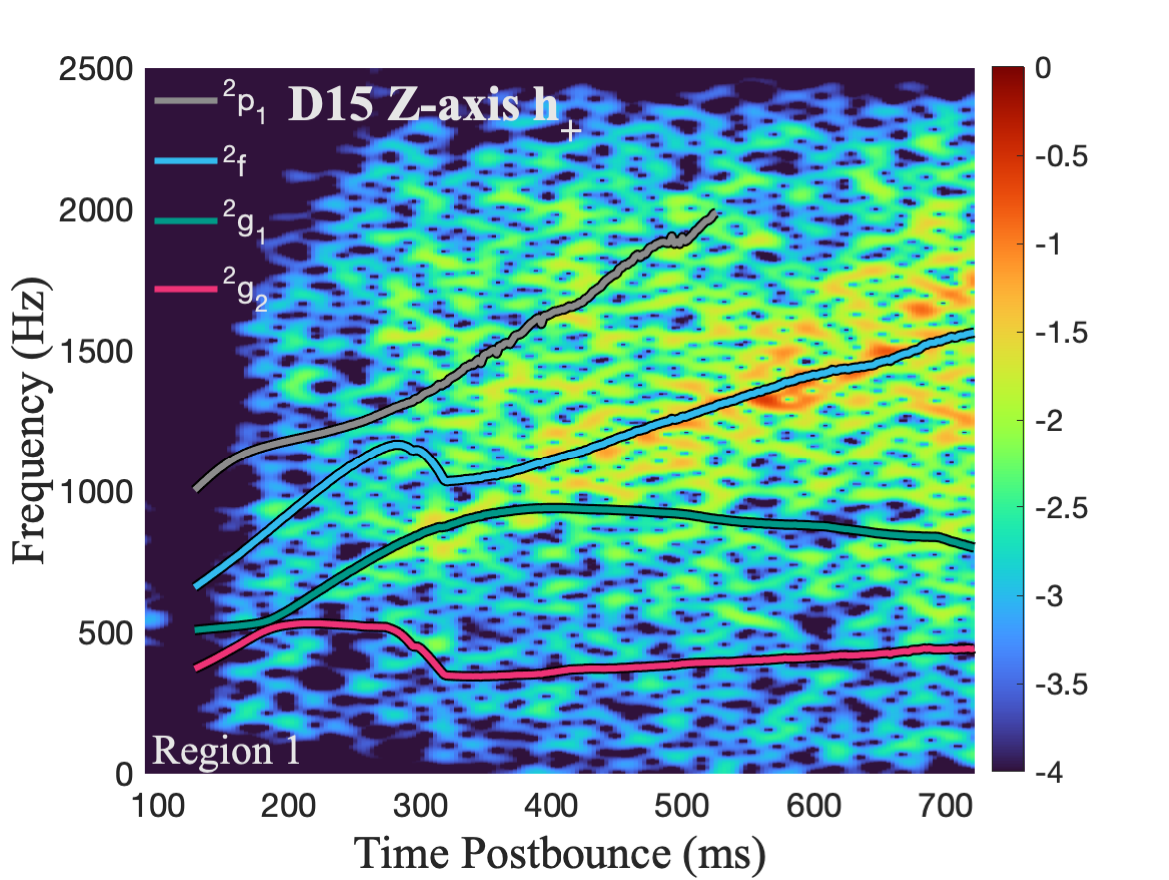}\hfill
     \includegraphics[width=8.5cm]{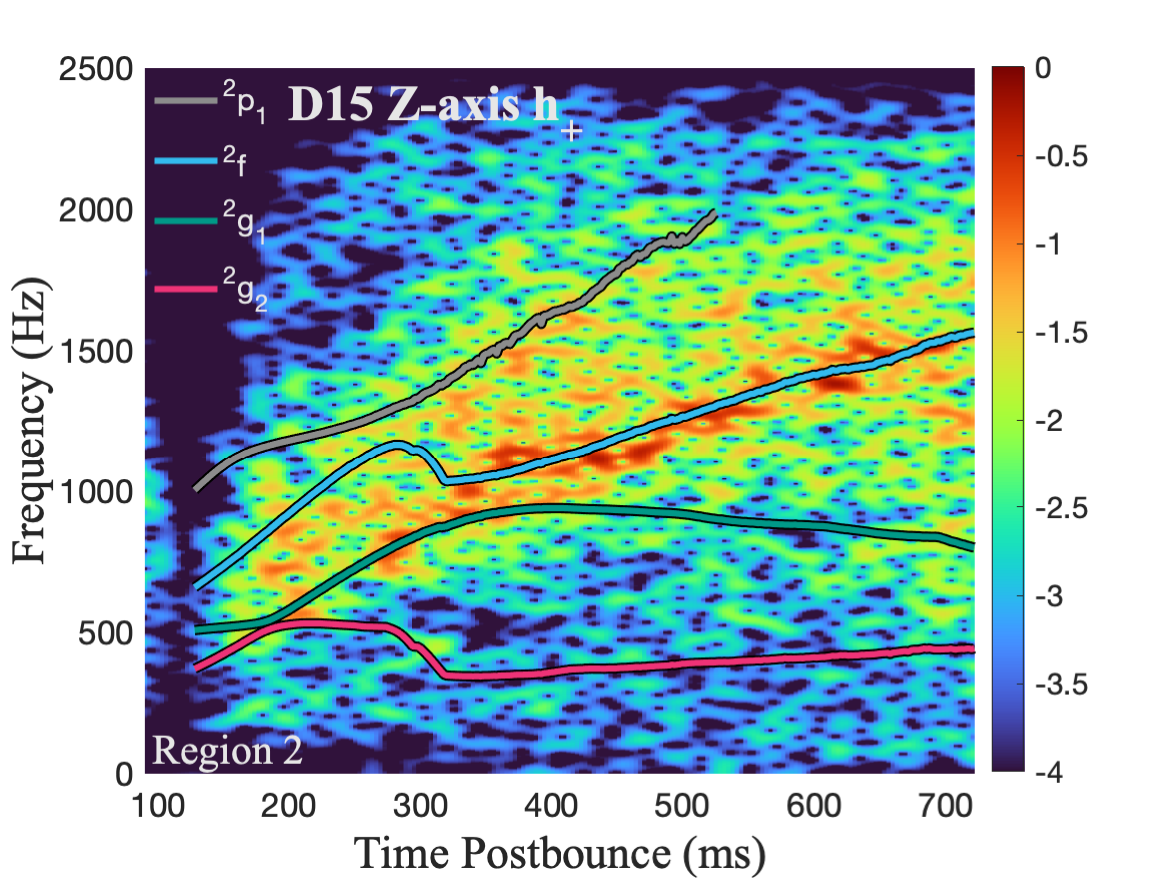}
     \includegraphics[width=8.5cm]{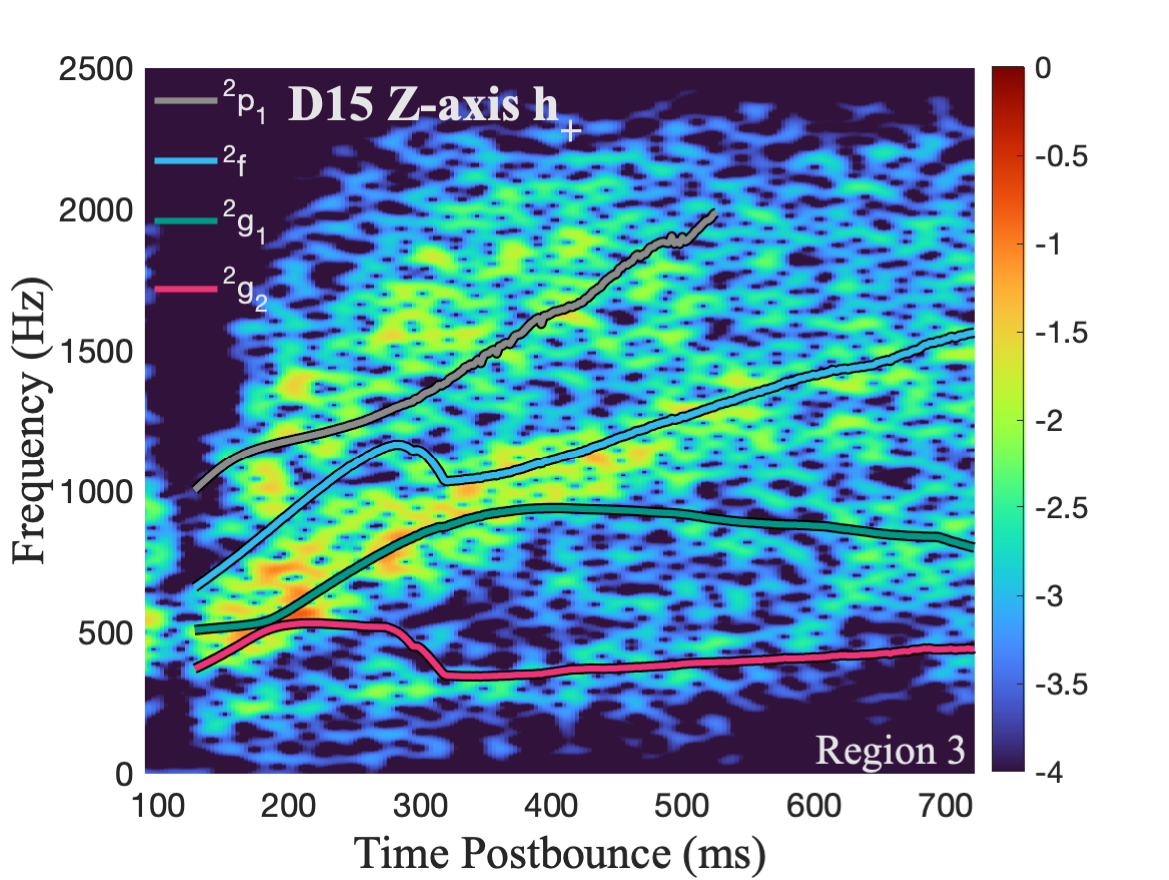}\hfill
     \includegraphics[width=8.5cm]{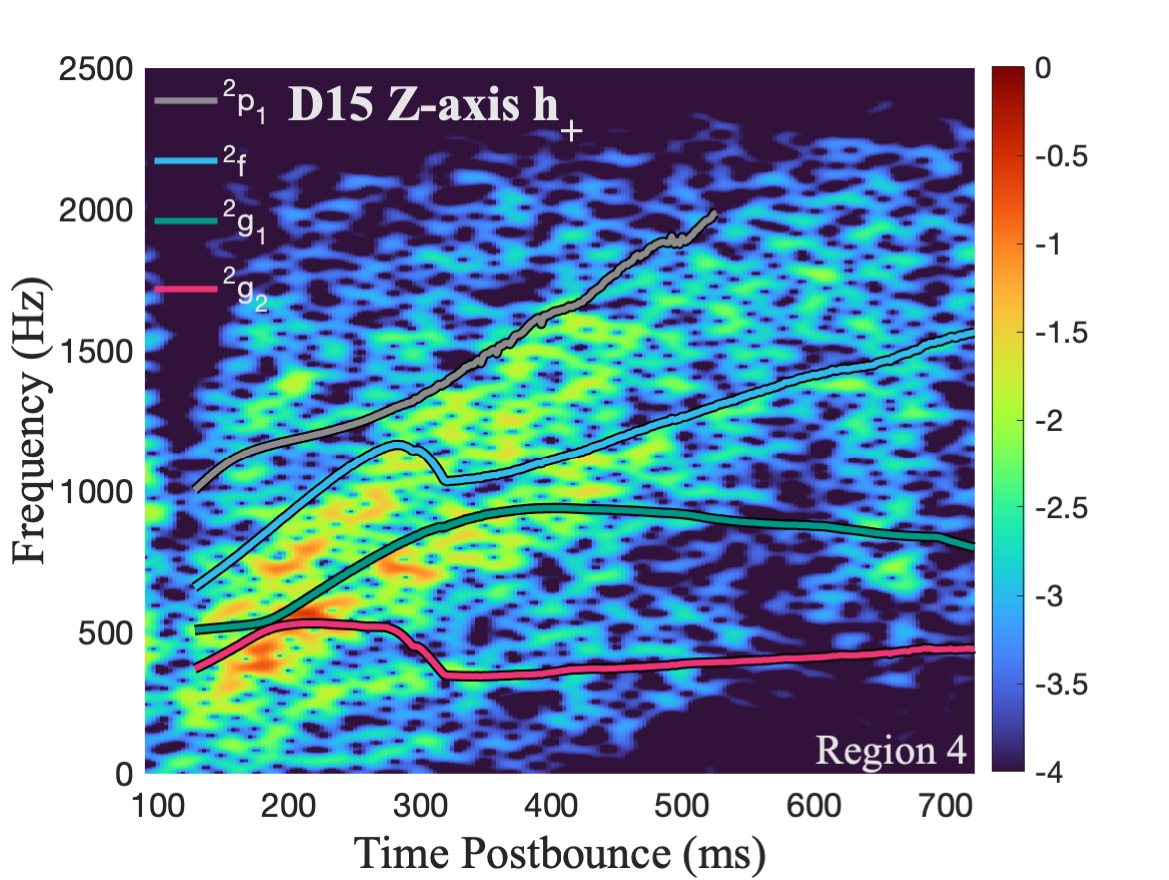}
     \caption{Spectrograms by region for model D15, corresponding to regional $h_+$ polarized strains in Figure \ref{fig:regional_strains_15}. The eigenfrequencies from the modal analysis are overlaid and differentiated by color. Modes are classified according to the GCN classification, as described in the text.}
     \label{fig:D15_rs}
\end{figure*}
\begin{figure*}
    \centering
     \includegraphics[width=8.5cm]{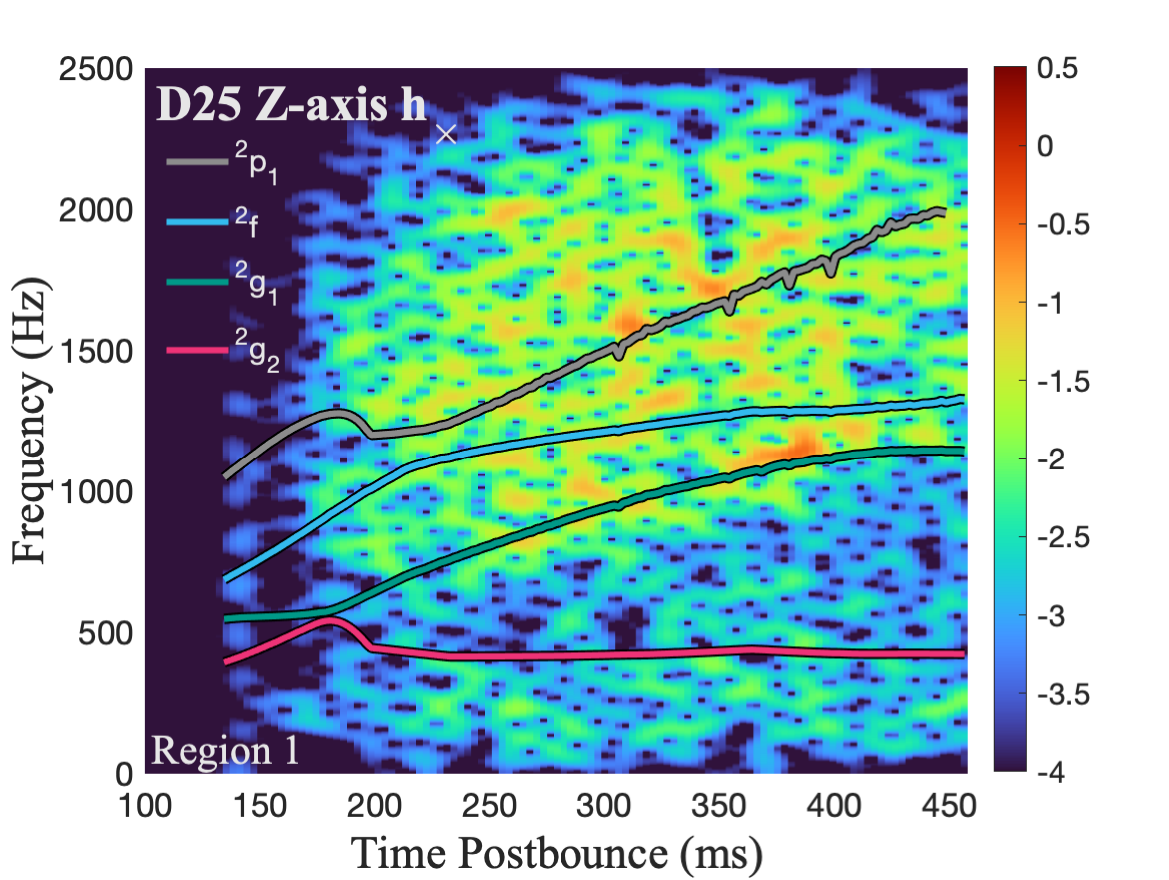}\hfill
     \includegraphics[width=8.5cm]{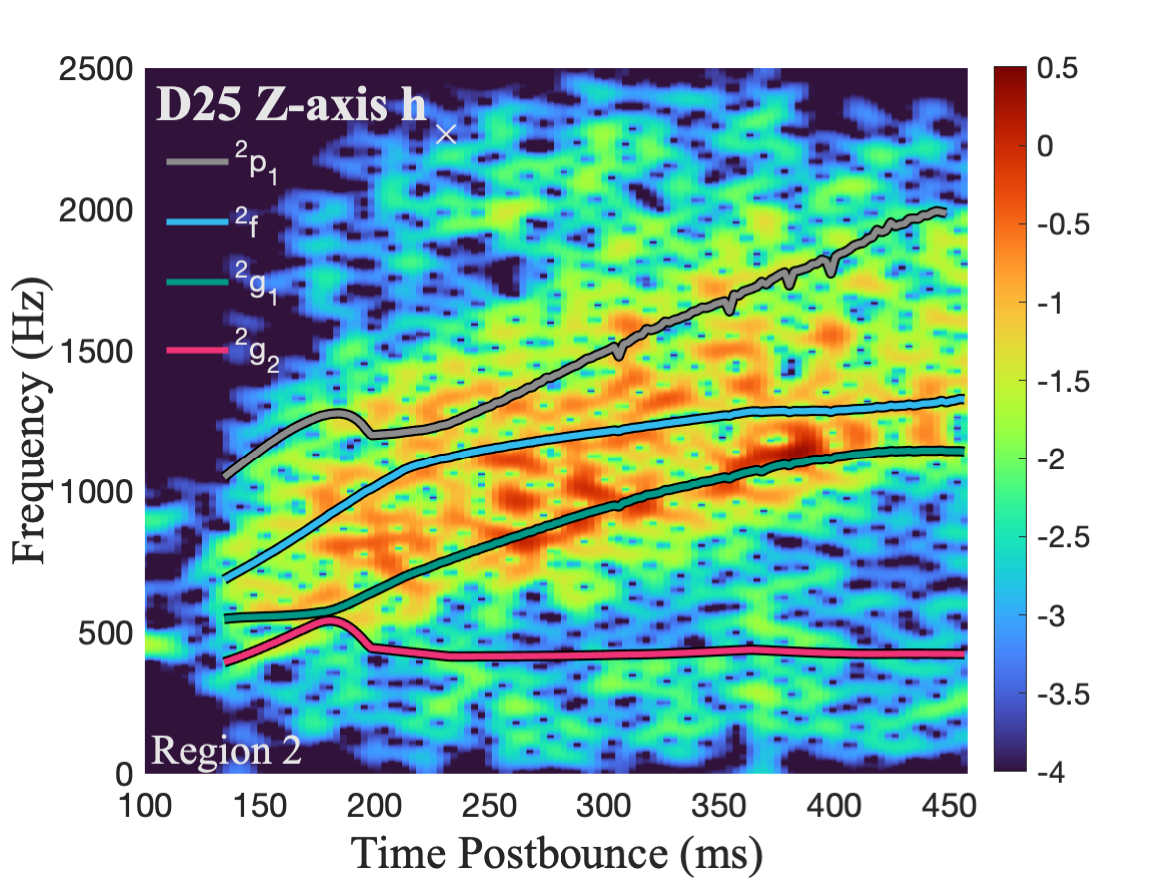}
     \includegraphics[width=8.5cm]{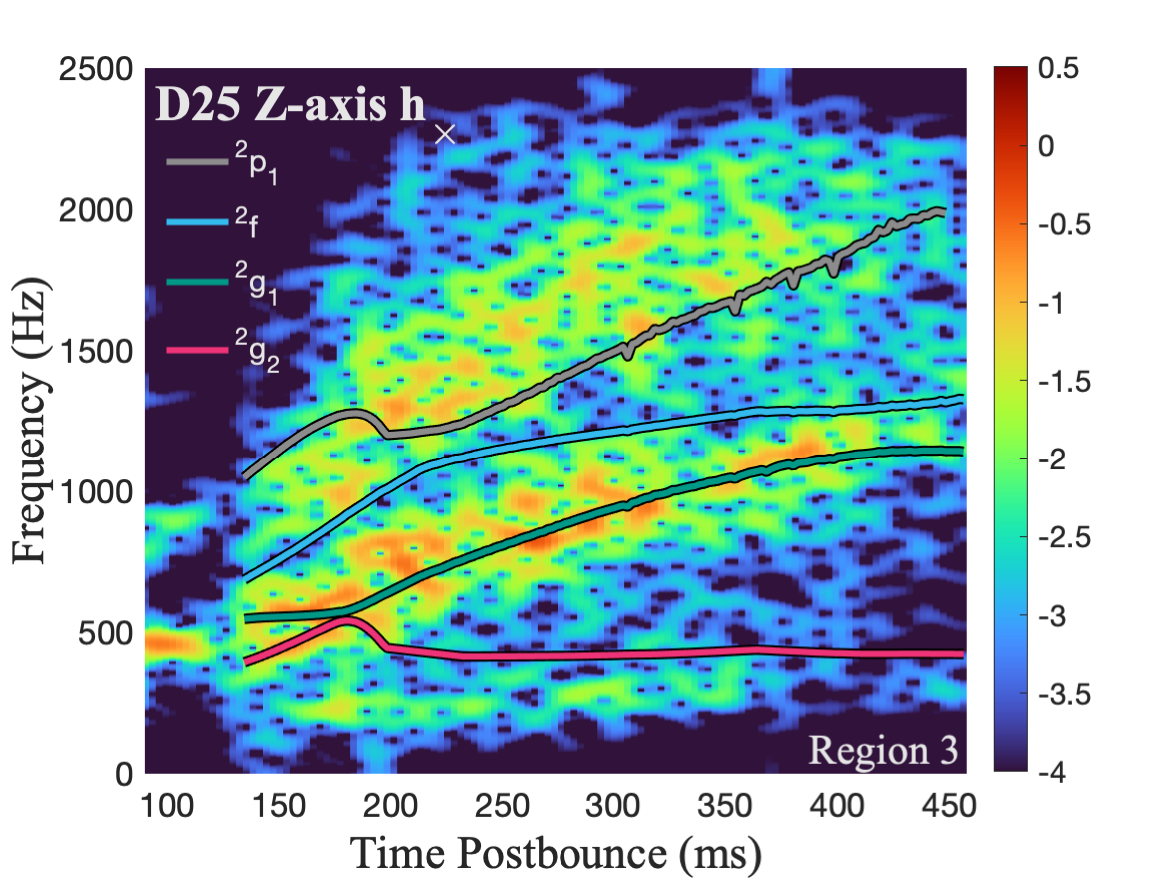}\hfill
     \includegraphics[width=8.5cm]{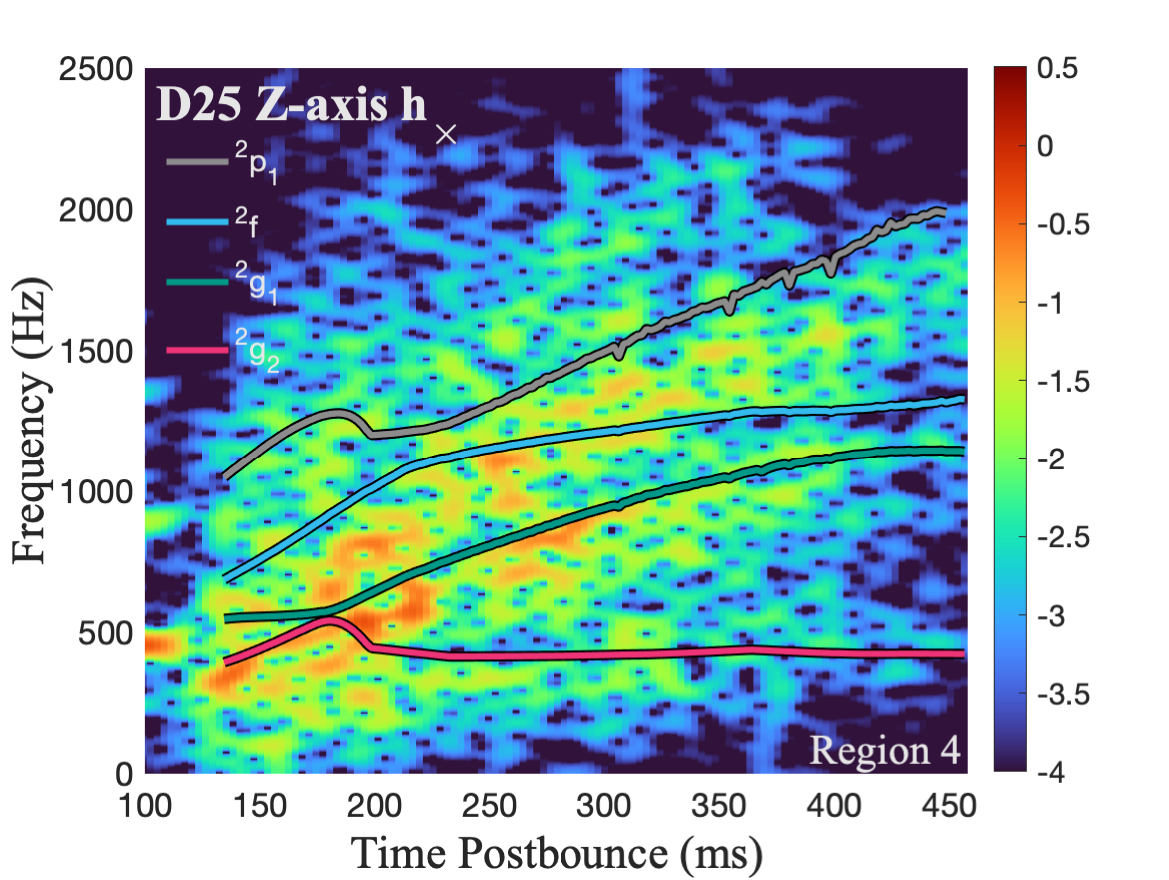}
     \caption{Spectrograms by region for model D25, corresponding to regional $h_\times$ polarized strains in Figure \ref{fig:regional_strains_25}. The eigenfrequencies from the modal analysis are overlaid and differentiated by color. Modes are classified according to the GCN classification, as described in the text.}
     \label{fig:D25_rs}
\end{figure*}

Finally, we examine the individual spectrograms for regions 1--4, given the corresponding strains in Figures \ref{fig:regional_strains_15} and \ref{fig:regional_strains_25}. Figures \ref{fig:D15_rs} and \ref{fig:D25_rs} show these spectrograms for model D15 and D25, respectively, produced using the same parameters as in Figure \ref{fig:Mode_comp}, overlaid with the eigenfrequencies that lie closest to the HFF from our modal analysis. The evolution of the source of gravitational wave emission is seen in the spectrograms and directly corroborated by the power emitted by each mode in Figure \ref{fig:Mode_power}. In both the D15 and D25 models, gravitational wave emission begins with the highest amplitude strains coming from accretion onto the PNS, in regions 3 and 4, corresponding to low-order ${}^2g$-modes. At these early times, there is no feature in region 1 that is tracked by these modes. Conversely, in the D15 model, after $\sim$300 ms postbounce, the ${}^2f$-mode is clearly tracked in region 2, and at later times in region 1, as well. During the same period, there is a weaker signal in region 3, and almost no signal in region 4. The D25 model does not evolve long enough for the ${}^2f$-mode to track the HFF, but the ${}^2g_1$-mode does track the HFF in both regions 2 and 3, with the dominant contribution coming from region 2 and with little or no contribution coming from region 3 or 4, respectively, towards the end of the simulation.

\section{Conclusions}
\label{sec:summary}
We have presented the first analysis of CCSN gravitational wave emission, using data from sophisticated three-dimensional CCSN simulations, that combines spatial decomposition of gravitational wave emission with a modal analysis of the emission, in an effort to source and characterize it. For the spatial analysis, we have introduced some improvements to the approach detailed in \citet{MeMaLa23}. For the modal analysis, we use a new tool based on a modal analysis appropriate for simulation data generated by a pseudo-Newtonian approach to gravity, consistent with the treatment of gravity implemented in \chimera. The necessity of the latter was demonstrated by \citet{West20}. 

\subsection{Summary of Results}
For the spatial decomposition of the gravitational wave emission, we computed, as in \cite{MeMaLa23}, the gravitational wave strains from five separate regions: (i) the region of sustained Ledoux convection deep within the PNS, at densities above $10^{12}$ g cm$^{-3}$ (region 1), (ii) the region of convective overshoot (region 2), (iii) the surface layer of the PNS, below a density of $10^{11}$ g cm$^{-3}$, which we use to define the PNS surface (region 3), (iv) the net neutrino cooling layer between the PNS surface and the gain radius (region 4), and finally, (v) the region above the gain radius (region 5). For the two models considered here, the general evolution of the gravitational wave strain amplitudes by region is well described as follows: (i) The strains start with small amplitudes in region 4 at $\sim$100 ms postbounce. (ii) The strain amplitudes rise to similar levels in region 3 by $\sim$110 ms postbounce. (iii) The strain amplitudes of region 2 become comparable to those of regions 3 and 4 by $\sim$160 ms postbounce. (iv) Simultaneously, the strains in region 1 also increase but remain slightly below those of regions 2, 3, and 4. (v) The strains in regions 1 and 2 continue to grow while, starting at $\sim$440 ms postbounce for D15 and $\sim$300 ms postbounce for D25, the strains in regions 3 and 4 begin to decrease. (vi) The strains in region 1 and 2 become approximately constant, on average, with the highest amplitude strains. 

In addition to computing the gravitational wave strains by region and, in particular, following the evolution of the strain amplitudes, we computed the fractional gravitational wave luminosities for our two models, as a function of radius from the center of the PNS. The strains are decomposable by region and can be added to give the total strain. The regional luminosities cannot be added to obtain the total value, but its approximate localization can be accomplished by integrating outward in radius. We find in both of our models that the evolution of the radius within which 90--95\% of the luminosity produced is consistent with the evolution of the largest strains, with the 90--95\% luminosity contours moving inward with postbounce time from regions 3 and 4 to regions 1 and 2.

The modal analysis, which complements the spatial analysis, as shown in Figures \ref{fig:Mode_power}, \ref{fig:D15_rs} and \ref{fig:D25_rs}, together providing a more complete picture, affords a third window onto the sources of gravitational emission. We computed a measure of the power emitted as gravitational waves in each region. At the beginning of the HFF, the ${}^2g_2$-mode is the most powerful, neglecting the early evolution of the ${}^2f$-mode, and region 3 dominates. (Recall that our modal analysis does not extend past region 3.) By $\sim$180 ms postbounce, the ${}^2g_1$-mode becomes dominant, and regions 2 and 3 radiate gravitational waves with approximately the same power. Over the next $\sim$30 ms, by $\sim$210--220 ms postbounce, region 2 becomes the dominant production region of gravitational waves. By $\sim$360 ms for D15 and $\sim$420 ms for D25, the ${}^2f$-mode has become the dominant producer of gravitational waves. In the D15 model, the dominant region remains region 2 until region 1 becomes slightly larger at the end of the simulation. For D25, region 1 becomes dominant $\sim$330 ms and remains dominant for the rest of the simulation.

Thus, all three approaches to sourcing the gravitational wave emission in the models presented here combine to elucidate our understanding of gravitational waves more completely and, moreover, tell a consistent story: Gravitational wave emission is dominated early by the PNS surface layers, as material accretes onto the PNS surface during the development of explosion, and dominated later by the region of sustained Ledoux convection and the region of convective overshoot, deep within the PNS, as explosion develops and accretion onto the PNS surface decreases.

\subsection{Comparison with Other Work}

We focused our modal analysis on the prominent, canonical HFF present in our spectrograms and the spectrograms from other modeling groups. To label our eigenfrequencies, we used the GCN classification scheme. The modes that track the HFF in our two models are the ${}^2g_2$-mode at the onset of the HFF, the ${}^2g_1$-mode beginning at $\sim$180 ms postbounce, and the ${}^2f$-mode at $\sim$360 ms postbounce for D15, and at $\sim$420 ms postbounce for D25. Just as the HFF itself is highly sensitive to progenitor parameters (\textit{e.g.}, the effects of varying mass, demonstrated in \cite{AnMuMu17,VaBuRa19b}), this sensitivity is also present in the transition times, or avoided crossing times, observed in our two models. This evolution of the HFF from low-order ${}^2g_n$-modes to the ${}^2f$-mode is consistent with the findings from three-dimensional models in \citet{RaMoBu19} and \citet{NaTaKo22}, as well as the findings from two-dimensional models in \cite{SoTa20,MoRaBu18,ZhAnOc24,SoMuTa24}.

Using {\em indirect} rather than direct methods in the context of three-dimensional models, others have concluded that gravitational wave emission from the PNS is excited primarily from outside \cite{RaMoBu19,PoMu19,VaBuWa23}. In particular, \citet{RaMoBu19} calculate the flux of the accreted turbulent kinetic energy at the surface of the PNS and show a trend between it and the total gravitational wave energy emitted, across a range of progenitors from 9--60~\msun. \citet{PoMu19} find that this empirical relationship between the turbulent kinetic energy flux onto the PNS and the emitted gravitational wave energy fits the emission from their models, as well, for an ultra-stripped 3.5~\msun\, helium core and an 18~\msun\ progenitor. Additionally, \citet{PoMu19} also show that the non-radial kinetic energy within the convective zone of the PNS in their models grows over the same time period that the {\em total} gravitational wave strain amplitudes are decreasing. 
With regard to the latter, we {\em too} see our total strains decrease with time, as we have shown, as the contributions from the surface layers {\em decrease} and the contributions from the convective and convective-overshoot layers are more or less sustained over the later periods of our runs. Therefore, a decrease in the {\em total} strain does {\em not} imply that the primary excitation mechanism of gravitational wave emission from the PNS is accretion onto it {\em at all times}.
A similar analysis was conducted by \citet{VaBuWa23}, who correlated turbulent accretion onto the PNS with periods of maximum strain. Again, as we have shown here, we {\em expect} the period of maximum strain to correspond to the period in which {\em all} regions we have defined here contribute significantly to gravitational wave emission---i.e., when we have significant surface emission, excited by accretion onto the PNS, {\em and} significant emission from the convective and convective-overshoot layers, excited by sustained Ledoux convection. The total strain does not, and cannot, tell us how the strain is {\em partitioned} over the regions of the PNS. Therefore, it cannot tell us that the strains in the convective and convective-overshoot regions remain large after the total strain has dropped off as accretion-induced strains drop off.
Our results, based on {\em direct} methods, show that the origin of the high-frequency gravitational wave emission in core collapse supernovae, which stems from the PNS, is generally more complex than has been assessed by other methods, and time-dependent.

Finally, in the modal analysis by \citet{ZhAnOc24}, based on two-dimensional axisymmetric CCSN simulation data, they find that the same eigenmodes we have identified here as best fitting the HFF in our spectrograms also fit the HFF in their spectrograms, namely the ${}^2g_2$-, ${}^2g_1$-, and ${}^2f$-modes, with early times after bounce matched by ${}^2g_n$-modes with more nodes, up to $n=4$. They emphasize that the eigenfunctions that best match the HFF are {\em global}---i.e., that the amplitudes of the eigenfunctions are significant throughout the PNS. This global nature is also noted in the perturbative analysis of axisymmetric models in \citet{EgZhSc21}. Our results show the same global character, as seen directly in the amplitudes of the eigenfunctions plotted in Figures \ref{fig:g12_avoid} and \ref{fig:gf_avoid}, which are appreciably nonzero throughout the PNS. However, it is clear there exist times when the power output across these regions differs by at least an order of magnitude, especially beyond 300 ms postbounce, between regions 1 or 2 and region 3 as shown by Figure \ref{fig:Mode_power}. That is, while global in nature, the dominant contributions to gravitational wave emission can still be sourced.

\subsection{Lessons Learned}

The use of a joint analysis in this work, combining spatial and modal analyses, served an additional, important purpose. In particular, the remarkable consistency between the evolution of the regional strains and the evolution of the eigenfunctions associated with the modes responsible for the HFF in our spectrograms demonstrates the efficacy of {\em each} of the analyses individually. More specifically, as the largest strain amplitudes transitioned from being found in the surface layers of the PNS to being found in the interior regions of the PNS, there was a commensurate transition in the eigenfunctions themselves---i.e., in the PNS oscillation modes giving rise to the gravitational wave emission. In other words, {\em there was a fundamental change in the character of the source of the gravitational wave emission.} Thus, our modal analysis supports our spatial analysis, and vice versa. 
And the results of our joint analysis also emphasize the need to distinguish between the {\em driving} mechanism behind gravitational wave emission, the region from which the emission is largest at any given time, and the PNS oscillation mode associated with it. To wit, late-time high-frequency emission in both models discussed here is driven by sustained Ledoux convection, resulting in convective overshoot in a convectively-stable region in which oscillations are buoyancy-driven, resulting in $f$-mode, as labeled in GCN, oscillations of the PNS.

It is also important to note that both of the analyses used here are {\em independent of} the ambiguities associated with the {\em classification of} the PNS oscillation modes giving rise to the gravitational wave emission, as $f$-modes, $g$-modes, or $p$-modes. While we used the GCN classification to label modes, the Cowling classification scheme would label the early part of the ${}^2f$-mode differently because it has nodes during that time. Likewise, the third method investigated in \citet{RoRaCh23} would yet again result in different labels on some of our eigenmodes at some times. Nonetheless, the conclusions we draw here are based on the eigenfunctions and eigenfrequencies {\em themselves}, which are independent of their classification. Such classification remains the subject of active research. Its importance is in part due to the desire to use a CCSN gravitational wave detection for PNS parameter estimation---specifically, to extract the mass and/or radius of the PNS as it evolves during the postbounce epoch. Together with the association made in astroseismology of $g$-modes with surface modes, depending on surface gravity and evolving as $M_{\rm PNS}/R_{\rm PNS}^{2}$, and $f$-modes with fundamental modes, depending on mean density and evolving as $\sqrt{M_{\rm PNS}/R_{\rm PNS}^3}$, a correspondence between the HFF and $M_{\rm PNS}/R_{\rm PNS}^2$ or $\sqrt{M_{\rm PNS}/R_{\rm PNS}^3}$ can be made. Thus, knowing which $(M_{\rm PNS},R_{\rm PNS})$ relationship pertains to which period of the postbounce evolution would enable the extraction of PNS parameters (or at least functions of them) from fits to the HFF. The fact that different classification schemes lead to different $(M_{\rm PNS},R_{\rm PNS})$-relation assignments clouds our ability to achieve the goal of using a CCSN gravitational wave detection for PNS parameter estimation. However, none of this clouds our ability to source the gravitational wave emission and to associate with it specific eigenfunctions/eigenfrequencies of PNS oscillation.

\smallskip

\section*{Acknowledgements}
The authors would like to thank Viktoriya Morozova, David Radice, Colter Richardson, and Ryan Westernacher-Schneider for most helpful discussions. 

A.M. acknowledges support from the National Science Foundation's Gravitational Physics Theory Program through grants PHY-1806692, PHY-2110177, and PHY-2409148. P.M. is supported by the National Science Foundation through its employee IR/D program. The opinions and conclusions expressed herein are those of the authors and do not represent the National Science Foundation.

This research used resources of the Oak Ridge Leadership Computing Facility at the Oak Ridge National Laboratory, which is supported by the Office of Science of the U.S. Department of Energy under Contract No. DE-AC05-00OR22725. 

\bibliography{add_journals,pr_add_journals,apj_journals,clean}

\end{document}